\documentclass{aastex63}
\usepackage{multirow}
\usepackage{graphicx}
\usepackage{mwe}
\usepackage[utf8]{inputenc}
\usepackage{xcolor}
\usepackage{xspace}
\usepackage{comment}
\usepackage{placeins}
\usepackage[version=4]{mhchem}

\newcommand{\firstedit}[1]{\textcolor{black}{#1\xspace}}
\newcommand{\response}[1]{\textcolor{black}{#1\xspace}}
\newcommand{\responsetwo}[1]{\textcolor{black}{#1\xspace}}

\newcommand{\zeratio}{$6.1\pm2.4$\xspace}

\received{June 1, 2019}
\revised{January 10, 2019}
\accepted{\today}
\submitjournal{AJ}

\shorttitle{Isomers in Interstellar Environments (I)}
\shortauthors{Shingledecker et al.}
\graphicspath{{./}{figures/}}

\begin{document}

\title{Isomers in Interstellar Environments (I): The Case of Z- and E-Cyanomethanimine}

\correspondingauthor{Christopher N. Shingledecker}
\email{cns@mpe.mpg.de}

\author[0000-0002-5171-7568]{Christopher N. Shingledecker}
\affil{Center for Astrochemical Studies
Max Planck Intitute for Extraterrestrial Physics
Garching, Germany}
\affil{Institute for Theoretical Chemistry 
University of Stuttgart 
Pfaffenwaldring 55, 70569 
Stuttgart, Germany}
\affil{Department of Physics \& Astronomy,
Benedictine College,
Atchison, Kansas 66002, USA}

\author[0000-0001-8803-8684]{Germ\'an Molpeceres}
\affil{Institute for Theoretical Chemistry 
University of Stuttgart 
Pfaffenwaldring 55, 70569 
Stuttgart, Germany}

\author[0000-0002-2887-5859]{V\'ictor M. Rivilla}
\affil{INAF-Osservatorio Astrofisico di Arcetri
Largo E. Fermi, 5
I-50125, Firenze, Italy}

\author[0000-0001-7031-8039]{Liton Majumdar}
\affil{School of Earth and Planetary Sciences, National Institute of Science Education and Research, HBNI, Jatni 752050, Odisha, India}

\author[0000-0001-6178-7669]{Johannes K\"astner}
\affil{Institute for Theoretical Chemistry 
University of Stuttgart 
Pfaffenwaldring 55, 70569 
Stuttgart, Germany}



\begin{abstract}

In this work, we present the results of our investigation into the chemistry of Z- and E-cyanomethanimine (\ce{HNCHCN}), both of which are possible precursors to the nucleobase adenine. Ab initio quantum chemical calculations for a number of reactions with atomic hydrogen were carried out. We find that the reaction H + Z/E-HNCHCN leading both to H-addition as well as H$_2$-abstraction proceed via similar short-range barriers with bimolecular rate coefficients on the order of $\sim10^{-17}$ cm$^{3}$ s$^{-1}$. These results were then incorporated into astrochemical models and used in simulations of the  giant molecular cloud G+0.693. The calculated abundances obtained from these models were compared with previous observational data and found to be in good agreement, with a predicted [Z/E] ratio of $\sim3$ - somewhat smaller than the previously derived value of $6.1\pm2.4$\xspace. We find that the [Z/E] ratio in our simulations is due mostly to ion-molecule destruction rates driven by the different permanent dipoles of the two conformers. Based on these results, we propose a general rule-of-thumb for estimating the abundances of isomers in interstellar environments. 

\end{abstract}

\keywords{editorials, notices --- 
miscellaneous --- catalogs --- surveys}


\section{Introduction} \label{sec:intro}

\response{Isomers, i.e. molecules with identical chemical formul\ae\xspace but differing structural arrangements of the constituent atoms, represent an intriguing subset of known interstellar species. From the perspective of astronomy, they are of interest, in part, since they can serve as useful indicators of the physical conditions of interstellar environments. For example, as was recently shown by \citet{hacar_hcn--hnc_2020}, the relative intensities of the $J=1-0$ line of hydrogen cyanide (HCN) and its isomer hydrogen isocyanide (HNC) can be used as a probe of the kinetic temperature of the gas. Moreover, isomers are also of interest from a chemical perspective since, again using HCN and HNC as an example, the metastable form is sometimes observed to be the most abundant \citep{irvine_cyanide_1984,schilke_study_1992,ungerechts_chemical_1997}, an unlikely situation under terrerstrial conditions.}

The HCN dimer, C-cyanomethanimine (\ce{H2C2N2}), is a molecule of particular astrochemical interest, given it's possible role as precursor to more complex prebiotic species such as adenine (\ce{H5C5N5}) \citep{oro_mechanism_1961,chakrabarti_can_2000}. That such a synthetic pathway could occur under astrophysical conditions has been suggested by the detection of a rich array of nucleobases - the monomeric building-blocks of RNA and DNA - in meteorites (see, e.g. \citet{ehrenfreund_extraterrestrial_2001} and \citet{callahan_carbonaceous_2011}). 

Two isomers of C-cyanomethanimine are the Z (trans) and E (cis) conformers \citep{clemmons_possible_1983}. Of these two, the Z-isomer is the more stable by 370 K, with the energy barrier between isomerization between the two being quite high, at 15.95 kK \citep{zaleski_detection_2013}. The Z- and E-isomers further differ in their permanent dipoles of $\mu_\mathrm{tot}^Z = 1.41$ D and $\mu_\mathrm{tot}^E = 4.11$ D, as measured by \citet{takano_microwave_1990}, with values of $\mu_\mathrm{tot}^Z = 0.62$ D and $\mu_\mathrm{tot}^E = 3.91$ D having been earlier calculated by \citet{clemmons_possible_1983}.  The larger permanent dipole of the E-isomer likely played a role in the fact that it was the first of the two conformers to be detected in the ISM by \citet{zaleski_detection_2013} using data from the Green Bank Telescope (GBT) PRIMOS survey of Sgr B2(N)\footnote{\url{https://www.cv.nrao.edu/PRIMOS/}}. Despite also searching for the Z-isomer, Zaleski and coworkers reported a non-detection of that less polar species. The Z form, together with the E, was later observed by one of us towards the giant molecular cloud G+0.693 \firstedit{in data from an IRAM 30m  spectral survey \citep{rivilla_abundant_2019}. This \response{molecule-rich} region near the Galactic Center has previously been studied in a number of works \citep{guesten_temperature_1985,huettemeister_kinetic_1993,rodriguez-fernandez_warm_2001,ginsburg_dense_2016,krieger_survey_2017}, and has been found to exhibit a particularly rich nitrile chemistry \citep{zeng_complex_2018}.}

Based on the observational data, a [Z/E] ratio of \zeratio was derived. \response{This fairly high abundance ratio was an intriguing result,} since existing branching ratios for known formation routes should yield values of only [Z/E]$\sim$1.5 in the case of the gas-phase reaction

\begin{equation}
    \ce{CN + CH2NH -> Z/E-HNCHCN + H}
    \label{R1}
    \tag{R1}
\end{equation}

\noindent
as studied by \citet{vazart_cyanomethanimine_2015}, or [Z/E]$\sim$0.9 in the gas of the grain-surface route

\begin{equation}
    \ce{NCCN + 2H -> Z/E-HNCHCN}
    \label{R2}
    \tag{R2}
\end{equation}

\noindent
which has been investigated by \citet{shivani_formation_2017}.

\firstedit{So then, the question remains: what gives rise to the [Z/E] ratio in G+0.693? It has been proposed that the relative abundances of isomers in interstellar environments could be estimated \textit{a priori} based on their thermodynamic stabilities \citep{lattelais_interstellar_2009,lattelais_new_2010}. Using the relative stabilities derived experimentally by \citet{takano_microwave_1990}, we previously estimated what temperature would be necessary to yield the observed [Z/E] using the expression}

\begin{equation}
    \left[\mathrm{Z}/\mathrm{E}\right] = \frac{N(\mathrm{Z})}{N(\mathrm{E})} = \frac{1}{g}\times \mathrm{exp}\left(\frac{\Delta E}{T_\mathrm{k}} \right)
    \label{thermo}
\end{equation}

\noindent
\firstedit{where here, $\Delta E$ is the difference in energy of the two conformers, and $g$ is a factor - here equal to 1 - that accounts for statistical weights. From Eq. \eqref{thermo}, a [Z/E]$\approx$6 implies a kinetic gas temperature of 130 - 210 K, which does overlap with the range of temperatures previously derived for G+0.693. However, under interstellar conditions, one is unlikely to ever obtain such a thermodynamic equilibrium and, as shown in part by \citet{herbst_calculations_2000}, \citet{loomis_investigating_2015}, and \citet{shingledecker_case_2019}, the relative abundances of isomers in interstellar environments are mostly kinetically determined, and therefore, a knowledge of key reaction barriers is critical is trying to make sense of observational results. However, it is often unclear which chemical processes give rise to the relative abundances of interstellar isomers.}

In a few cases, e.g. for ion-neutral reactions, there are typically no short-range barriers (activation energies), only long-range centrifugal ones which the translational energy of the reactants is sufficient to overcome \citep{herbst_gas_2006}. In such cases, a satisfactory upper limit to the reaction rate-coefficient can be estimated using capture theory \citep{woon_quantum_2009}. For non-polar molecules, the rate coefficient in this case is given by the Langevin formula, 

\begin{equation}
    k_\mathrm{L} = 2\pi e \sqrt{\frac{\alpha}{\mu}}
\end{equation}

\noindent
where here $e$ is the electronic charge and $k_\mathrm{L}$, the Langevin rate, is a function of the dipole polarizability, $\alpha$, and the reduced mass of the reactants, $\mu$. In the case of reactions between ions and neutral molecules with permanent dipoles, models typically use the expression of \citet{su_parametrization_1982}, which accounts for the enhanced long-range attraction and is given by

\begin{equation}
   k_\mathrm{D} = k_\mathrm{L}(0.4767x + 0.6200) 
   \label{kdlow}
\end{equation}

\noindent
or

\begin{equation}
    k_\mathrm{D} = k_\mathrm{L}\left[ \frac{(x + 0.5090)^2}{10.526} + 0.9754 \right]
    \label{kdhigh}
\end{equation}

\noindent
with $x$ being expressed as

\begin{equation}
    x = \frac{\mu_\mathrm{D}}{\sqrt{2\alpha k_\mathrm{B}T}},
    \label{kdx}
\end{equation}

\noindent
where $k_\mathrm{B}$ is the Boltzmann constant. In the approach of Su and Chesnavich, Eq. \eqref{kdlow} is used in cases where $x \geq 2$, and Eq. \eqref{kdhigh} when $x < 2$, though when $x=0$, $k_\mathrm{D} = k_\mathrm{L}$. 

Conversely, for most bimolecular reactions involving neutral species, there usually exist some short-range activation energy barriers, in which case determining the rate coefficients is usually non-trivial. Reactions between radicals and closed-shell neutral molecules, which can be barrierless, represent an occasional exception to this general rule-of-thumb. For interstellar chemistry, perhaps the most important radical is atomic hydrogen, which is efficiently produced in even dense molecular clouds from the dissociation of \ce{H2} \citep{padovani_production_2018}. However, even isomers that are fairly structurally similar can have markedly different \response{reactivities} with H (and other radicals), as we have previously shown with proadienone and propynal, two isomers in the \ce{H2C3O} family of molecules \citep{shingledecker_case_2019}. This surprising result was used by us to help explain why propadienone, despite being the most stable of the \ce{H2C3O} species \citep{karton_pinning_2014}, had consistantly eluded detection \citep{loomis_investigating_2015,loison_interstellar_2016}, thereby illustrating the importance of reactions with atomic hydrogen - and of destruction processes, more generally - in understanding interstellar isomer abundances. 

Thus, in order to better understand the physicochemical mechanisms leading to the [Z/E]$=$\zeratio reported in \citet{rivilla_abundant_2019}, we have carried out an investigation of the reactivity of atomic hydrogen with both Z/E-cyanomethanimine and several related species. The rest of this work is organized as follows: in \S \ref{sec:methods} we describe the method used in our calculations, the results of which are summarized in \S \ref{sec:results}. In \S \ref{sec:modeling} we investigate the effects of the reactions we have studied under astrophysical conditions. Finally, our conclusions are given in \S \ref{sec:conclusions}.

\section{Methods} \label{sec:methods}

In this work, we have \response{examined} two different mechanisms \response{which} could, in principle, account for the observed abundance of the Z isomer  \response{by means of \textit{ab initio} quantum chemical calculations.} \response{Specifically, we have studied} the oxidation/reduction of \ce{HNCHCN} via reactions with H atoms.

\begin{figure}[ht]
\begin{center}
\includegraphics[width=0.4\textwidth]{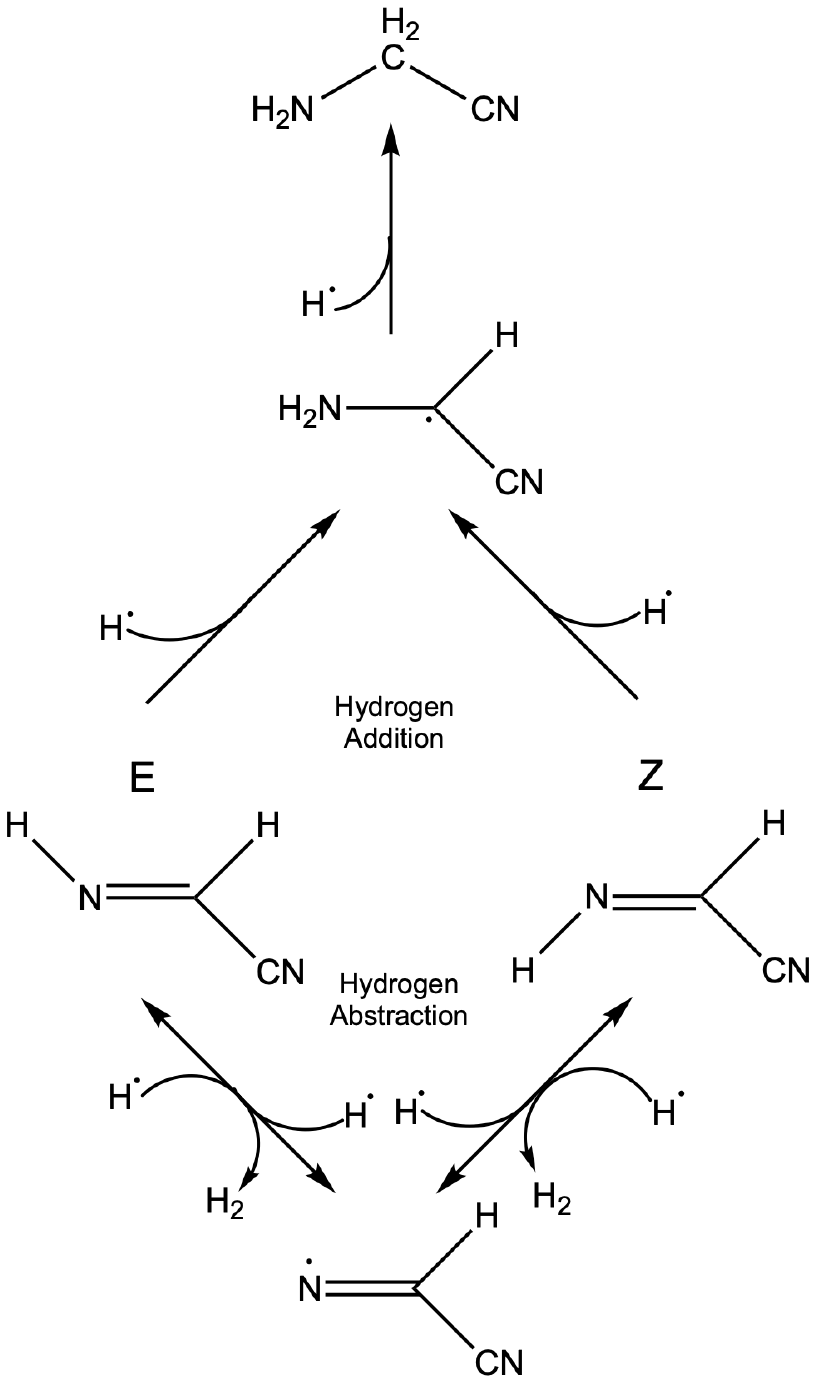}
\caption{Reaction network simulated in this work.}
\label{scheme_reactions}
\end{center}
\end{figure}

\response{To this end,} we have constructed a partial reaction network \response{consisting of} two different types of reactions \response{for} each isomer, namely, hydrogen addition on dust grains and molecular hydrogen abstraction. From the parent species, these reactions lead to the formation of a reactive radical that can subsequently \response{react again with H, thereby returning} to the closed-shell neutral form or, in the case of two consecutive hydrogen additions, to amino-acetonitrile, a positively identified molecule in the interstellar medium \citep{Belloche2008}, thought to be relevant in the formation of interstellar glycine \citep{Koch2008,Danger2011,Kolesnikova2017} . Fig \ref{scheme_reactions} shows a schematic view of this network.

For the study of each of the individual reactions we have characterized the stationary points on \response{their} respective potential energy \response{surfaces} by means of density functional theory (DFT) calculations. We have employed the mPWB1K exchange \& correlation functional \response{of} Zhao \& Truhlar \citep{Zhao2004} in combination with the def2-TZVP basis set \citep{Weigend2005}. The \response{main} advantage of this functional \response{for this investigation} is that \response{it was} designed \response{for the easy} determination of activation energies. Minima and transition states (TS) were optimized using the DL-FIND \citep{Kastner2009} program of the ChemShell suite \citep{Sherwood2003,Metz2014}. A search of TS was done by means of potential energy surface scans and posterior optimization in cases were a kinetic barrier was predicted. Assessment of the nature of the stationary points was performed by computing the molecular Hessian \response{of} each. \responsetwo{Additionally, we have addressed the validity of the DFT method by computing single-point calculations on the relevant DFT geometries at the \responsetwo{CCSD(T)-F12/cc-PVTZ-F12} level of theory using Molpro 2015 \citep{Molpro, molpro2015}. We have found excellent agreement between DFT and coupled cluster methods for the reaction barriers, with deviations of less than 1 kcal/mol in all cases}. Furthermore, we have carried out intrinsic reaction coordinate (IRC) calculations to ensure proper connection between our calculated TS and their associated minima. For reactions with a barrier, we have computed \response{both} classical bimolecular reaction rate coefficients as well as tunneling-corrected ones. The inclusion of quantum tunneling in our calculations has been performed using semi-classical instanton theory \citep{Rommel2011,Rommel2011-2}, following a sequential cooling scheme for temperatures below the crossover temperature, and reduced instanton theory for temperatures above \citep{McConnell2017}. Crossover temperatures are defined - with $\nu_{i}$ being the frequency of the vibrational imaginary mode in the transition state, and k$_{B}$ the Boltzmann constant - as:

\begin{equation}
    T_{c} = \frac{\hbar \nu_{i}}{k_{B}} .
\end{equation}

For T $<$ T$_{c}$ we started with an even distribution of images at a temperature T$_{ini}\sim$ 0.7 T$_{c}$ and finished at a temperature of 50 K. All the electronic structure calculations were carried out using the Gaussian09 suite of programs (revision D.01) \citep{g09}.

\section{Results and Discussion} \label{sec:results}
\subsection{Chemical processing with H: Addition and Abstraction Reactions}

\begin{table}[ht]
\caption{Abbreviated names used in discussions of our quantum chemical calculations.}
\label{tab:nomenclature}
\centering
\begin{tabular}{cc}
\hline
\hline
Species & Abbreviation \\
\hline
\ce{Z-HNCHCN} & R-Z \\
\ce{E-HNCHCN} & R-E \\
\ce{H2NCHCN} & P1 \\
\ce{NCHCN} & P2 \\
\hline
\end{tabular}
\end{table}

\begin{deluxetable*}{ccllllcc}[bt!]
\tablecaption{Reactions for Z- and E-cyanomethanimine added to our network. \label{tab:reactionTable}}
\tablehead{
\colhead{Number}   &
\colhead{Label}    &
\colhead{Reaction} &
\colhead{$\alpha$} &
\colhead{$\beta$}  &
\colhead{$\gamma$} &
\colhead{Formula}  &
\colhead{Phase}
}
\startdata
1  & R1.1 & \ce{CN + CH2NH -> E-HNCHCN + H}          & $3.2\times10^{-10}$ & $2.0\times10^{-1}$ & $9.1\times10^{-2}$ & \textit{a}         & \textit{gas}$^*$   \\
2  & R1.2 & \ce{CN + CH2NH -> Z-HNCHCN + H}          & $5.0\times10^{-10}$ & $2.0\times10^{-1}$ & $8.7\times10^{-2}$ & \textit{a}         & \textit{gas}$^{*}$   \\
3  & R2.1 & \ce{HNCCN + H -> E-HNCHCN}               & $5.0\times10^{-1}$  & $0.0\times10^{0}$  & $0.0\times10^{0}$  & \textit{b}         & \textit{grain}$^{**}$ \\
4  & R2.2 & \ce{HNCCN + H -> Z-HNCHCN}               & $5.0\times10^{-1}$  & $0.0\times10^{0}$  & $0.0\times10^{0}$  & \textit{b}         & \textit{grain}$^{**}$ \\
5  & R3.1 & \ce{H + E-HNCHCN -> H2NCHCN}             & $1.0\times10^{0}$   & $0.0\times10^{0}$  & $1.4\times10^{3}$  & \textit{b}         & \textit{grain} \\
6  & R3.2 & \ce{H + Z-HNCHCN -> H2NCHCN}             & $1.0\times10^{0}$   & $0.0\times10^{0}$  & $1.3\times10^{3}$  & \textit{b}         & \textit{grain} \\
7  & R3.3 & \ce{H + E-HNCHCN -> NH3 + CCN}           & $9.8\times10^{-12}$ & $3.3\times10^{0}$  & $9.5\times10^{2}$  & \eqref{kinstanton} & \textit{gas}$^\dagger$   \\
8  & R3.4 & \ce{H + Z-HNCHCN -> NH3 + CCN}           & $1.5\times10^{-11}$ & $3.3\times10^{0}$  & $9.8\times10^{2}$  & \eqref{kinstanton} & \textit{gas}   \\
9 & R4.1 & \ce{H2NCHCN + H -> H3NCHCN}              & $1.0\times10^{0}$   & $0.0\times10^{0}$  & $0.0\times10^{0}$  & \textit{b}         & \textit{grain} \\
10  & R4.2 & \ce{H2NCHCN + H -> H2 + E-HNCHCN}        & $1.5\times10^{-10}$ & $0.0\times10^{0}$  & $0.0\times10^{0}$  & \textit{a}         & \textit{gas}   \\
11 & R4.3 & \ce{H2NCHCN + H -> H2 + Z-HNCHCN}        & $1.5\times10^{-10}$ & $0.0\times10^{0}$  & $0.0\times10^{0}$  & \textit{a}         & \textit{gas}   \\
12 & R5.1 & \ce{H + E-HNCHCN -> NCHCN + H2}          & $3.8\times10^{-13}$ & $1.5\times10^{0}$  & $1.2\times10^{3}$  & \eqref{kinstanton} & \textit{gas}   \\
13 & R5.2 & \ce{H + Z-HNCHCN -> NCHCN + H2}          & $4.9\times10^{-14}$ & $2.8\times10^{0}$  & $9.5\times10^{2}$  & \eqref{kinstanton} & \textit{gas}   \\
14 & R5.3 & \ce{H + E-HNCHCN -> NCHCN + H2}          & $1.0\times10^{0}$   & $0.0\times10^{0}$  & $4.4\times10^{3}$  & \textit{b}         & \textit{grain}   \\
15 & R5.4 & \ce{H + Z-HNCHCN -> NCHCN + H2}          & $1.0\times10^{0}$   & $0.0\times10^{0}$  & $4.8\times10^{3}$  & \textit{b}         & \textit{grain}   \\
16 & R6.1 & \ce{NCHCN + H -> E-HNCHCN}               & $5.0\times10^{-1}$  & $0.0\times10^{0}$  & $0.0\times10^{0}$ & \textit{b}         & \textit{grain} \\
17 & R6.2 & \ce{NCHCN + H -> Z-HNCHCN}               & $5.0\times10^{-1}$  & $0.0\times10^{0}$  & $0.0\times10^{0}$ & \textit{b}         & \textit{grain} \\
18 & R6.3 & \ce{NCHCN + H -> H2 + C2N2}              & $3.0\times10^{-10}$ & $0.0\times10^{0}$  & $0.0\times10^{0}$ & \textit{a}         & \textit{gas}   \\
19 & -   & \ce{H3+ + E-HNCHCN -> H2 + H2CN + HNC+}  & $1.0\times10^{0}$   & $4.0\times10^{-9}$ & $7.2\times10^{0}$ & \eqref{kdlow}/\eqref{kdhigh} & \textit{gas}$^\ddagger$   \\
20 & -   & \ce{H3+ + Z-HNCHCN -> H2 + H2CN + HNC+}  & $1.0\times10^{0}$   & $4.0\times10^{-9}$ & $2.3\times10^{0}$ & \eqref{kdlow}/\eqref{kdhigh} & \textit{gas}   \\
21 & -   & \ce{H+  + E-HNCHCN -> H2 + CN + HNC+}    & $1.0\times10^{0}$   & $6.8\times10^{-9}$ & $7.2\times10^{0}$ & \eqref{kdlow}/\eqref{kdhigh} & \textit{gas}   \\
22 & -   & \ce{H+  + Z-HNCHCN -> H2 + CN + HNC+}    & $1.0\times10^{0}$   & $6.8\times10^{-9}$ & $2.3\times10^{0}$ & \eqref{kdlow}/\eqref{kdhigh} & \textit{gas}   \\
23 & -   & \ce{He+  + E-HNCHCN -> He + HCN + HNC+}  & $1.0\times10^{0}$   & $4.9\times10^{-9}$ & $7.2\times10^{0}$ & \eqref{kdlow}/\eqref{kdhigh} & \textit{gas}   \\
24 & -   & \ce{He+  + Z-HNCHCN -> He + HCN + HNC+}  & $1.0\times10^{0}$   & $4.9\times10^{-9}$ & $2.3\times10^{0}$ & \eqref{kdlow}/\eqref{kdhigh} & \textit{gas}   \\
25 & -   & \ce{C+ + E-HNCHCN -> C + HCN + HNC+}     & $1.0\times10^{0}$   & $2.2\times10^{-9}$ & $7.2\times10^{0}$ & \eqref{kdlow}/\eqref{kdhigh} & \textit{gas}   \\
26 & -   & \ce{C+ + Z-HNCHCN -> C + HCN + HNC+}     & $1.0\times10^{0}$   & $2.2\times10^{-9}$ & $2.3\times10^{0}$ & \eqref{kdlow}/\eqref{kdhigh} & \textit{gas}   \\
27 & -   & \ce{HCO+ + E-HNCHCN -> H2CO + CN + HNC+} & $1.0\times10^{0}$   & $1.6\times10^{-9}$ & $7.2\times10^{0}$ & \eqref{kdlow}/\eqref{kdhigh} & \textit{gas}   \\
28 & -   & \ce{HCO+ + Z-HNCHCN -> H2CO + CN + HNC+} & $1.0\times10^{0}$   & $1.6\times10^{-9}$ & $2.3\times10^{0}$ & \eqref{kdlow}/\eqref{kdhigh} & \textit{gas}   \\
29 & -   & \ce{H3O+ + E-HNCHCN -> H2O + H2CN + HNC+}& $1.0\times10^{0}$   & $1.8\times10^{-9}$ & $7.2\times10^{0}$ & \eqref{kdlow}/\eqref{kdhigh} & \textit{gas}   \\
30 & -   & \ce{H3O+ + Z-HNCHCN -> H2O + H2CN + HNC+}& $1.0\times10^{0}$   & $1.8\times10^{-9}$ & $2.3\times10^{0}$ & \eqref{kdlow}/\eqref{kdhigh} & \textit{gas}   \\
\enddata
\begin{minipage}{\textwidth}
	\footnotesize
	\vspace{1em}
	(a) $k = \alpha \left(\frac{T}{300\;\mathrm{K}}\right)^{\beta} \mathrm{exp}\left(-\frac{\gamma}{T}\right) \mathrm{cm^{3}\;s^{-1}}.$ \\
	(b) Grain-surface rate coefficients are calculated as described in \S2.3 of \citet{ruaud_gas_2016}. Here, $\alpha$ represents the unitless branching fraction and $\gamma$ is the value of the activation energy in Kelvin. \\
	(*) Taken from \citet{vazart_cyanomethanimine_2015} \\
	(**) Taken from \citet{shivani_formation_2017} \\
	($\dagger$) Assuming a value of $T_0=150$ K for all reactions using Eq. \eqref{kinstanton} \\
	($\ddagger$) The temperatures at which the model would switch from Eq. \eqref{kdlow} to \eqref{kdhigh} for all E and Z cyanomethanimine ion-neutral reactions are 4070 K and 480 K, respectively. Here, $\alpha$ is a unitless branching fraction, $\beta$ is the Langevin rate, and $\gamma$ is the value of Eq. \eqref{kdx} evaluated at $T=300$ K. \\
	\normalsize
\end{minipage}
\vspace{-5em}
\end{deluxetable*}

In order to check the possibility of a chemically-induced isomerization, we have simulated the reaction scheme presented in Fig \ref{scheme_reactions} \response{and corresponding to reactions R3 - R6 in Table \ref{tab:reactionTable}}. A brief comment on the nomenclature summarized in Table \ref{tab:nomenclature}: we will employ the notation Addition-X (X = E, Z) and H$_{2}$-abstraction-X (X = E, Z) through the text. With this we are referring to the position where the hydrogen atom acts. Therefore, the H-addition-E in \ce{NCHCN} will lead to the E-isomer and vice-versa. Reactants are named R-E and R-Z for the E and Z isomer, respectively. \response{The products of these processes we will call} P1, for \ce{H2NCHCN} (hydrogen addition), and P2, for \ce{NCHCN} (hydrogen abstraction).

We have computed the energy difference between E \& Z isomer to be 2.19 kJ/mol (263 K) at our level of theory, very close to the recent values of Puzzarini et al of 2.38 kJ/mol (286 K) at the CCSD(T) level of theory \citep{Puzzarini2015}.

\subsubsection{H-Addition to HNCHCN}

\response{We will first consider our  simulation of the reactions depicted in the upper portion of Fig. \ref{scheme_reactions}, i.e.}

\begin{equation}
    \ce{Z/E-HNCHCN + H -> products}.
    \label{R3}
    \tag{R3}
\end{equation}

\responsetwo{The grain-surface reactions} R-E + H $\rightarrow$ P1 (R3.1) and R-Z + H $\rightarrow$ P1 (R3.2) show an energy barrier of 14.36 kJ/mol (1727 K) and 13.96 kJ/mol (1679 K) (both ZPE corrected), respectively. Energies of the relevant structures for this reaction are shown in Table \ref{hadditionbarriers}.   

\begin{table}[ht]
\caption{Reaction energies and activation energies (in kJ/mol, with and without ZPE), vibrational frequencies of the TS imaginary mode, and crossover-temperatures T$_{c}$ for the atomic hydrogen addition reaction \eqref{R3} for both isomers considered in this work. \firstedit{In parenthesis, values employing CCSD(T)-F12.}}
\label{hadditionbarriers}
\centering
\begin{tabular}{c|cccccc}
\hline
\hline
Reaction & $\Delta$E$_R$ & $\Delta$E$_A$ &  $\Delta$U$^{0}_R$ & $\Delta$U$_{A}^{0}$ & T$_{c}$ (K) \\
 \hline
H+R-E $\rightarrow$ P1 & -247.61 (-226.85) & 11.23 (10.72) & -220.14 & 14.36 & 171 \\
H+R-Z $\rightarrow$ P1 & -245.42 (-224.41) & 10.97 (8.34) & -218.30 & 13.96 & 168\\
\hline
\end{tabular}
\end{table}

\begin{figure}[ht]
\begin{center}
\includegraphics[width=0.5\textwidth]{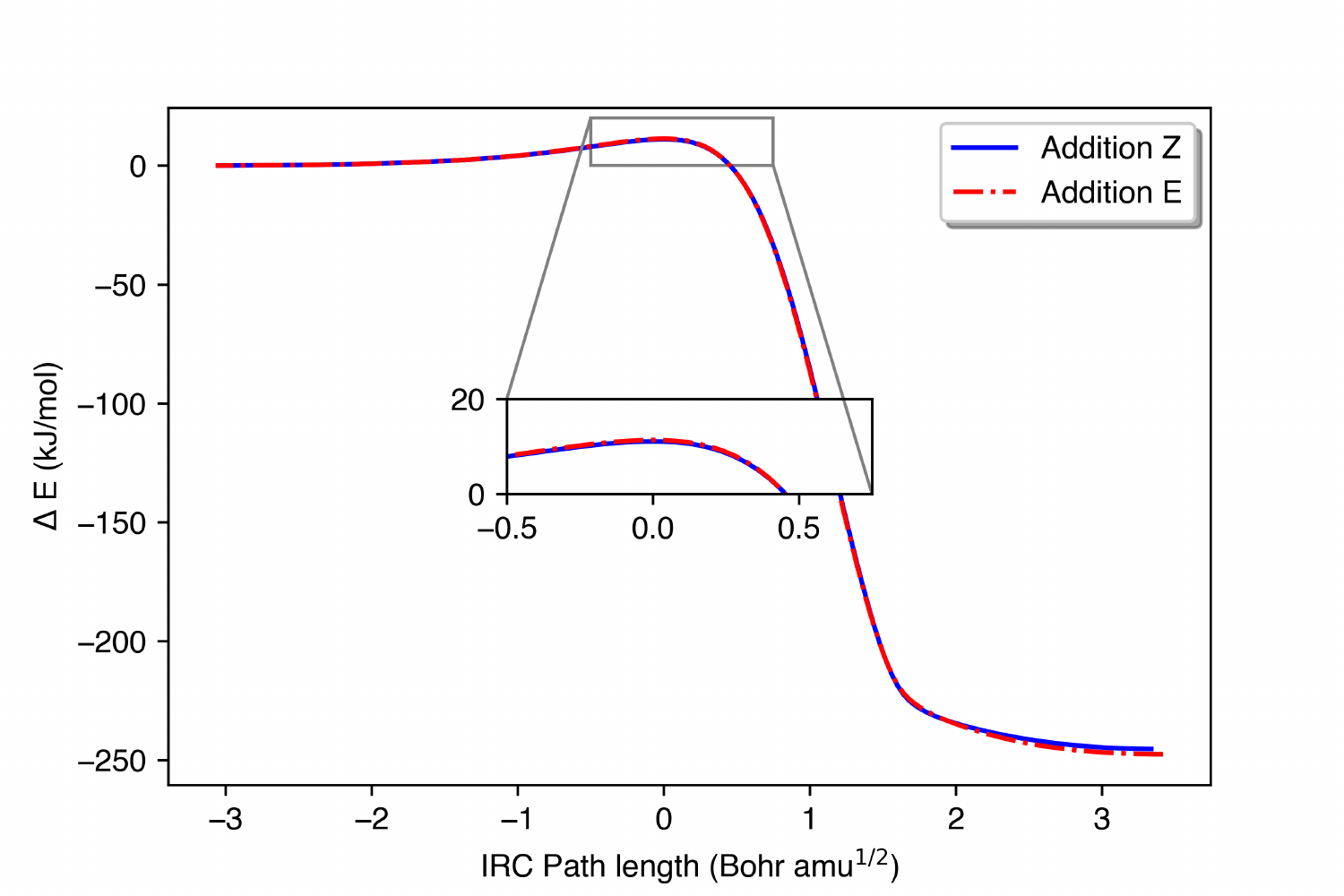}
\caption{IRC energy profile for the H-Addition reactions. Energies are not ZPE corrected.}
\label{barriersaddition}
\end{center}
\end{figure}

\response{Both reactions are exothermic and present similar activation energies}. The two reactions \response{thus} share almost the same profile, \response{a fact which is readily apparent from the plot of the IRC shown in Fig. \ref{barriersaddition}.}

\begin{figure}[ht]
\begin{center}
\includegraphics[width=0.5\textwidth]{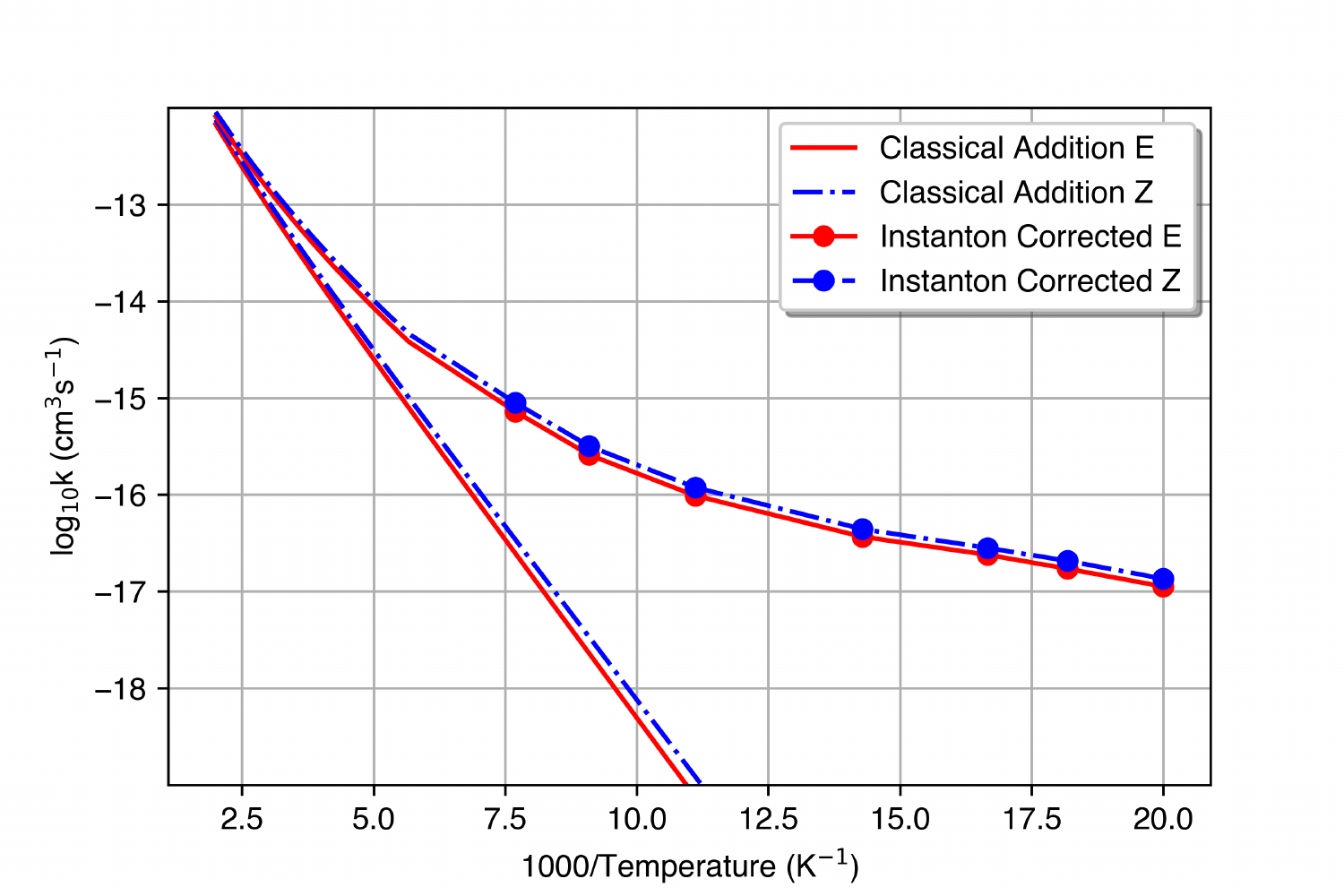}
\caption{Reaction rate constants for the \ce{HNCHCN} (E-Z) + H $\rightarrow$ \ce{H2NCHCN} reaction. Reduced instanton rate constants are represented as a continuous line above the crossover temperature. Lines connecting dots for instanton rate constants are a guide to the eye.}
\label{arrhenius_addition}
\end{center}
\end{figure}

\response{From Fig. \ref{barriersaddition}, it is obvious that both} profiles \responsetwo{are very similar}, with a slight difference in favor of the addition to the Z isomer. However, \response{these} differences are so small that they \response{can} hardly explain any preferential reactivity and thus, change in the abundances of the parent isomers. Arrhenius plots showing \responsetwo{gas-phase} reaction rate constants are presented in Fig \ref{arrhenius_addition}. Instanton corrected rate constants for this reaction are calculated using an instanton path of 20 images for temperatures above 50 K and using 58 images in the case of the rate at 50 K, for a better convergence of the path. Again, \response{differences in the magnitudes of these rate coefficients of less than 1\% are not sufficient in elucidatinig the underlying chemical mechanism responsible for the [Z/E]=\zeratio of \citet{rivilla_abundant_2019}}. \responsetwo{The calculated energy barriers for reaction \eqref{R3} were used in the determination of rate coefficients for the grain-surface form of the reaction, and the calculated rate constants shown in Fig. \ref{barriersaddition} were used for the gas-phase form of the reaction (see Table \ref{tab:reactionTable} and \S\ref{sec:modeling} for further details).}

\subsubsection{H addition to \ce{H2NCHCN}}

\response{We next turn our attention to the reaction}

\begin{equation}
    \ce{H2NCHCN + H -> products}
    \label{R4}
    \tag{R4}
\end{equation}

\response{which} is a fast, barrierless radical-radical recombination. From the possible \response{reaction} sites in \ce{H3C2N2}, the most likely is the central carbon. This assumption is also supported by the spin density at that atom of 0.627 a.u, as obtained from a Mulliken population analysis of the electronic wavefunction. \response{We have confirmed that \eqref{R4} presents a barrierless profile by checking a downhill path along the reaction coordinate}. \response{We find this reaction to be highly} exothermic, with a total reaction energy of \firstedit{$\Delta$E$_R$= -356.20 (-375.73) kJ/mol (in parenthesis, values using CCSD(T)-F12) and} $\Delta$U$^{0}_R$ = -319.71 kJ/mol.

\responsetwo{Such a barrierless reaction will occur every collision in the ISM, and so, as shown in Table \ref{tab:reactionTable}, we assume it occurs at the collisional rate of $3\times10^{-10}$ cm$^{3}$ s$^{-1}$, split equally among the E and Z product channels.}

\subsubsection{H$_{2}$-abstraction from \ce{HNCHCN}}

The second set of reactions that we have studied involves H$_{2}$ abstraction from \ce{HNCHCN} and \response{the subsequent} hydrogenation of the \response{resulting} radical. \response{Of these, we first consider} the abstraction reactions from (E,Z)- cyanomethinimine:

\begin{equation}
    \ce{Z/E-HNCHCN + H -> products.}
    \label{R5}
    \tag{R5}
\end{equation}

\responsetwo{For the grain-surface product channel} R-E + H $\rightarrow$ P2 (R5.1)  and  R-Z + H $\rightarrow$ P2 (R5.2) we proceeded in a similar \response{manner to our previously described metholodogy for the} additions,  \response{thereby} obtaining energetic barriers for both isomers. Energy values for both reactions are presented in Table \ref{h2eliminationbarriers}.

\begin{table}[ht]
\caption{Reaction energies and activation energies (in kJ/mol, with and without ZPE) and vibrational frequencies of the TS imaginary mode for the molecular hydrogen abstraction reaction \eqref{R5} for both isomers considered in this work. \firstedit{In parenthesis, values employing CCSD(T)-F12.}}
\label{h2eliminationbarriers}
\centering
\begin{tabular}{c|ccccc}
\hline
\hline
Reaction & $\Delta$E$_R$ & $\Delta$E$_A$ &  $\Delta$U$_{R}$$^{0}$ & $\Delta$U$_{A}$$^{0}$  & T$_{c}$ (K) \\
 \hline
H+R-E $\rightarrow$ P2 + H$_2$ & -38.29 (-39.63) & 36.59 (37.38)  & -48.68 & 28.49 & 420\\
H+R-Z $\rightarrow$ P2 + H$_2$ & -36.11 (-37.18) & 40.00 (39.83) & -46.84 & 31.58 & 428\\
\hline
\end{tabular}
\end{table}

\begin{figure}[ht]
\begin{center}
\includegraphics[width=0.5\textwidth]{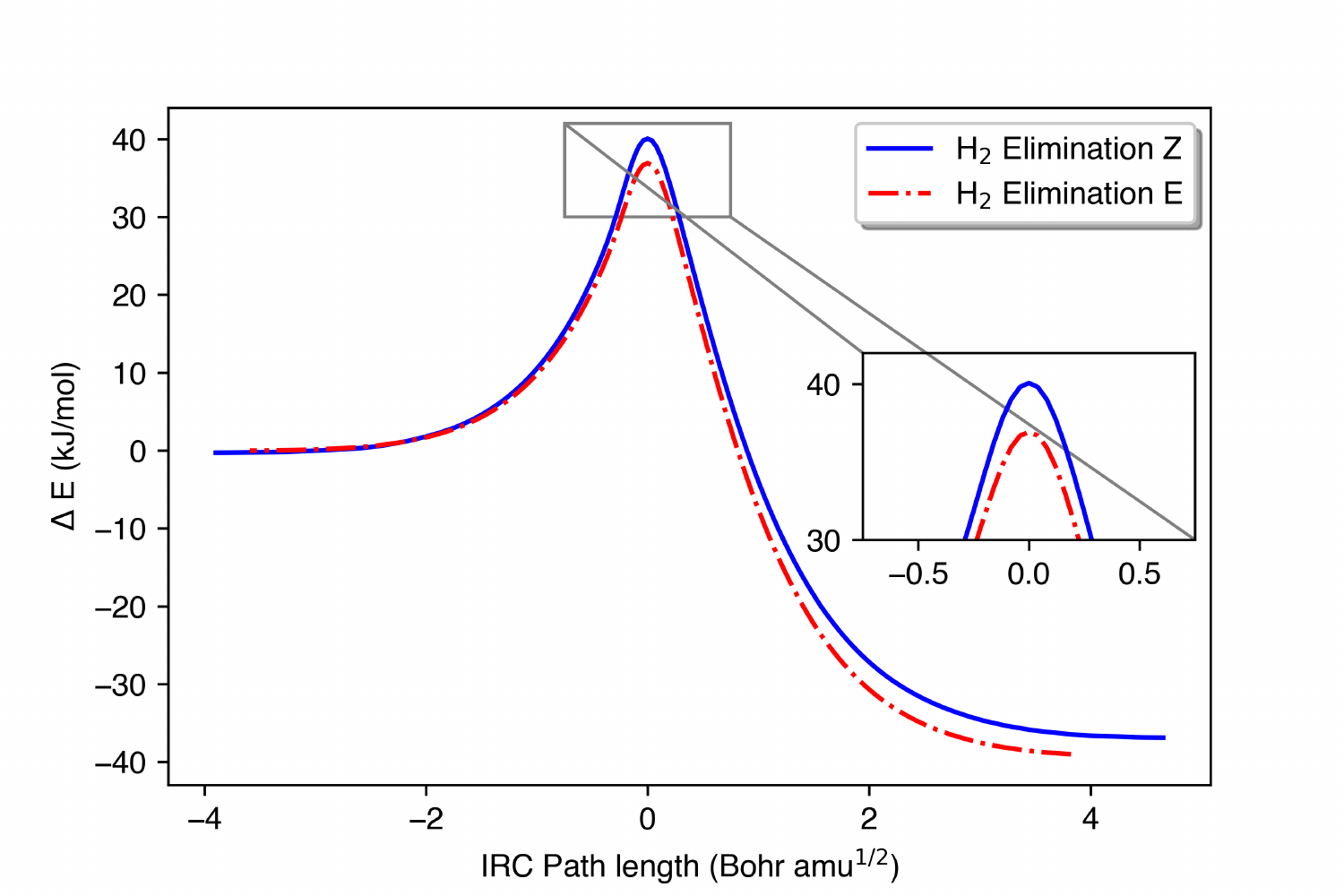}
\caption{IRC energy profile for the H$_{2}$-Abstraction reactions. \responsetwo{Energies are not vibrationally corrected}.}
\label{barrierselimination}
\end{center}
\end{figure}

This set of reactions is less exothermic than \eqref{R3} and proceeds with a higher activation barrier. However, in this case, the energetic separation between both transition states is larger, being somewhat higher in the case of the Z-isomer. This means that the H$_{2}$ abstraction reaction from the E isomer is more efficient than the same reaction involving the Z isomer. The shape of the barriers is obtained from computing the IRC profile in both reactions, as shown in Fig \ref{barrierselimination}.

From the profile we can see that the nature of the barrier is completely different than in the case of the H-Addition reaction. The differences in the profile between E and Z are also relevant, showing a wider one for the reaction to abstract \ce{H2} of the Z isomer. This has implications in the tunneling rate constants.

\begin{figure}[ht]
\begin{center}
\includegraphics[width=0.5\textwidth]{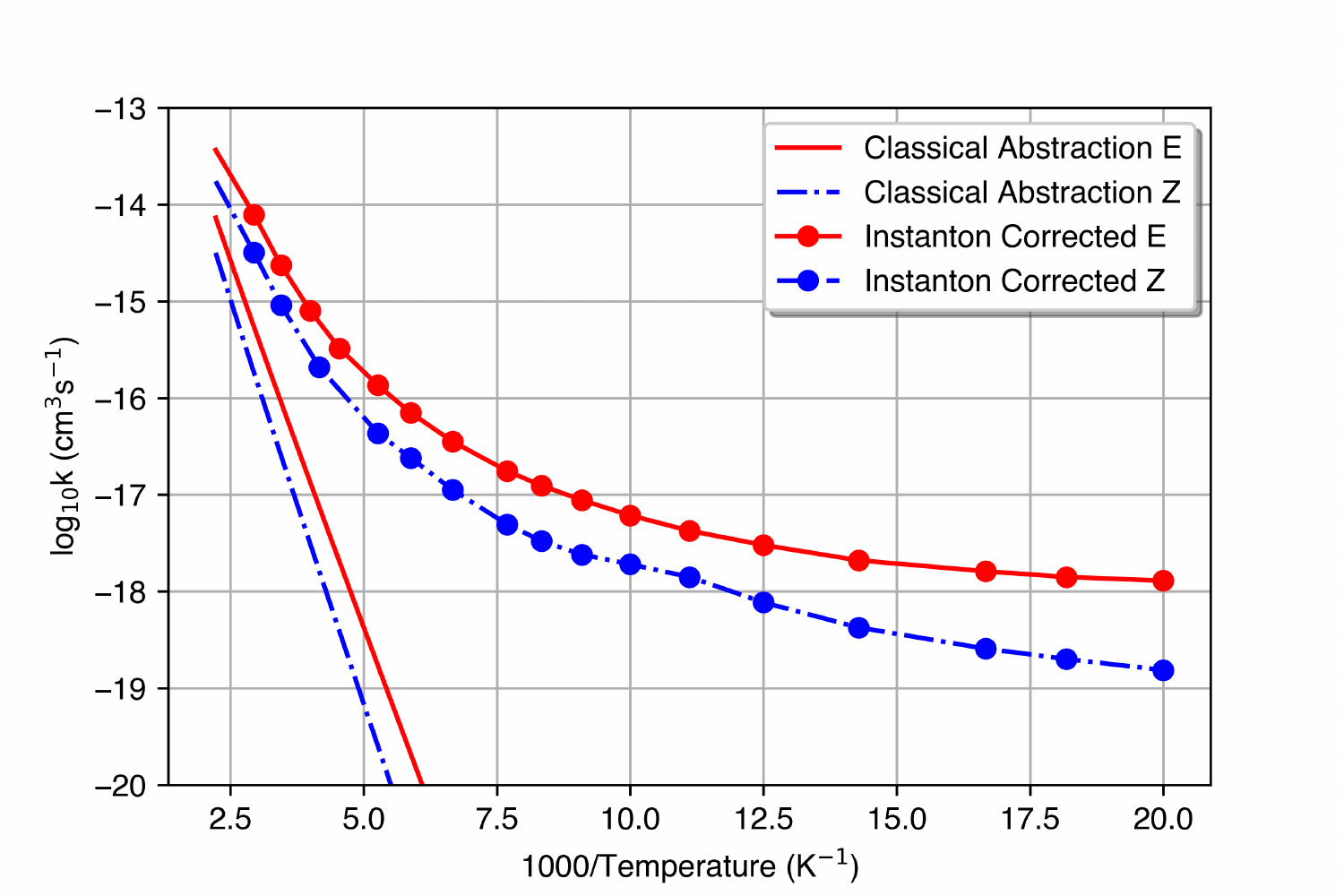}
\caption{Reaction rate constants for the H$_2$C$_{2}$N$_{2}$ (E-Z) + H $\rightarrow$ HC$_{2}$N$_{2}$ + H$_{2}$ reaction. Reduced instanton rate constants are represented as a continuous line above the crossover temperature. Lines connecting dots for instanton rate constants are a guide to the eye.}
\label{arrhenius_elimination}
\end{center}
\end{figure}

These hydrogen abstractions point to \responsetwo{the E isomer as being} preferentially destroyed \response{in these} \ce{H2}-abstraction reactions. In order to confirm \response{this supposition}, we have computed the tunneling-\response{corrected} bimolecular rate constants for both processes, shown in Fig \ref{arrhenius_elimination}.

From \response{Fig. \ref{arrhenius_elimination}} we can see that there is a gap between \responsetwo{the gas-phase rate coefficients} at 50 K, in favor of the E isomer, \response{indicating} that it reacts more easily with H at low temperatures. The ratio between both rate constants at 50 K is 8.38 in favor of the destruction of the E isomer. The separation between the rate constants becomes higher at even lower temperatures, so this behavior should become even more drastic below 50 K.

\responsetwo{As with reaction \eqref{R3}, the calculated energy barriers for reaction \eqref{R5} were used to determine rate coefficients for the grain-surface variant of the reaction, and the calculated rate constants shown in Fig. \ref{arrhenius_elimination} were used for the gas-phase form of the reaction. Further details can be found in Table \ref{tab:reactionTable} and \S\ref{sec:modeling}.}

\subsubsection{H-Addition to \ce{NCHCN} }

\response{Finally, we consider the case of hydrogen addition to the \ce{NCHCN} adduct:}

\begin{equation}
    \ce{NCHCN + H -> products}
    \label{R6}
    \tag{R6}
\end{equation}

This set of reactions, P2 + H $\rightarrow$ R-E (R6.1) \& P2 + H $\rightarrow$ R-Z (R6.1), can be also categorized as a radical recombination reaction and thus, a barrierless process without preferential stereochemistry.From the different atomic positions in which the addition can take place, the most likely is where the formal charge resides (see bottom part of Fig \ref{scheme_reactions}. This is confirmed also from the Mulliken population analysis of the optimized radical, with a spin density in the nitrogen where the formal charge resides of 1.05 a.u., illustrating that the unpaired electron is heavily localized on that atom.\response{The barrierless nature of \eqref{R6} has been checked as in \eqref{R4}.} ZPE corrected reaction energies are shown in Table \ref{proto_addition_energies}. Both reactions are very exothermic, with a slight difference between them due to the different stability of the product isomers. \responsetwo{Since this too is a barrierless process, as with reaction \eqref{R4}, we adopt the collisional rate of $3\times10^{-10}$ cm$^{3}$ s$^{-1}$, split equally among the E and Z product channels, for use in our chemical simulations.}

\begin{table}[ht]
\caption{Reaction energies (in kJ/mol, with and without ZPE) for the hydrogen addition reaction to the P$_{2}$ radical (HC$_{2}$N$_{2}$) \eqref{R6}.  \firstedit{In parenthesis, values employing CCSD(T)-F12.}}
\label{proto_addition_energies}
\centering
\begin{tabular}{c|cc}
\hline
\hline
Reaction & $\Delta$E$_R$ &  $\Delta$U$_{R}$$^{0}$ \\
 \hline
P2 + H $\rightarrow$ R-E & -408.99 (-419.56)  & -371.71 \\
P2 + H $\rightarrow$ R-Z & -411.17 (-422.00) & -373.55 \\
\hline
\end{tabular}
\end{table}

\subsubsection{Summary of the reactions}

\response{In this section} we have presented results concerning the reactions depicted in Fig \ref{scheme_reactions}. We find that the processing of cyanomethanimine with hydrogen atoms involves a series of slow processes. We have found a preferential reaction route to one or the other conformer in the case of the tunneling mediated hydrogen abstraction acting in the imine group at low temperatures, which favors the destruction of the E form. Furthermore, from the produced radical \ce{NCHCN}, further reaction with H is barrierless to reform (E,Z)-cyanomethanimine. This back and forth conversion between radicals thus predicts an excess in favor of the Z isomer, as suggested by the astronomical observations.

Another possibility for the reaction of the parent isomers with hydrogen is the addition of the atom to the imine group, to form an amino radical. The barriers for this process do not hint to any preferential stereochemistry. Moreover, a second hydrogen does not restore the previous two isomers, in contrast to the previous set of reactions. This second hydrogen is employed in the formation of amino acetonitrile. In any case, from this set of reactions no excess is predicted.

\response{As we have previously found with, e.g. reactions between atomic hydrogen and propadienone (\ce{H2C3O}), such radical/closed-shell neutral reactions can be very efficient under interstellar conditions and occur with rate coefficients of $\sim3\times10^{-10}$ cm$^3$ s$^{-1}$ \citep{shingledecker_case_2019}. However, for the reactions we have investigated here with non-zero activation energies, i.e. R3 and R5, the gas-phase rate coefficients shown in Figs. 3 \& 5 were found to be low ($\sim 10^{-17}-10^{-18}$ cm$^3$ s$^{-1}$) at temperatures relevant to molecular clouds. As we will describe in more detail in \S\ref{sec:modeling}, these results imply that the overall importance of these reactions will be small compared with more efficient processes, such as ion-neutral reactions, which have rate coefficients on the order of $\sim 10^{-9}$ cm$^3$ s$^{-1}$.}

\section{Astrochemical Modeling \& Implications} \label{sec:modeling}

\begin{deluxetable}{lc}[tb]
\tablecaption{Physical conditions utilized for results shown in Fig. \ref{fig:modelresults}. \label{tab:physparameters}}
\tablewidth{0pt}
\tablehead{
\colhead{Parameter} & \colhead{Value}   
}
\startdata
$n_\mathrm{gas}$ & $10^4$ cm$^{-3}$ \\
$A_\mathrm{V}$ & 10 mag\\
$\zeta$ & $1.3\times10^{-15}$ s$^{-1}$ \\ 
$T_\mathrm{g}$ & 150 K \\
$T_\mathrm{d}$ & 15 K \\
\enddata
\end{deluxetable}

\begin{deluxetable}{cc}[tb]
\tablecaption{Initial elemental abundances \label{tab:initialelemental}}
\tablewidth{0pt}
\tablehead{
\colhead{Element} & \colhead{Relative Abundance} 
}
\startdata
\ce{H2}  & $4.99\times10^{-1}$ \\
H        & $5.00\times10^{-5}$ \\
\ce{He}  & $9.00\times10^{-2}$ \\
O        & $2.40\times10^{-4}$ \\
\ce{C+}  & $1.40\times10^{-4}$ \\
N        & $6.20\times10^{-5}$ \\
\ce{Mg+} & $7.00\times10^{-9}$ \\
\ce{Si+} & $8.00\times10^{-9}$ \\
\ce{Fe+} & $3.00\times10^{-9}$ \\
\ce{S+}  & $8.00\times10^{-8}$ \\
\ce{Na+} & $2.00\times10^{-9}$ \\
\ce{Cl}  & $1.00\times10^{-9}$ \\
\ce{P+}  & $2.00\times10^{-10}$ \\
\ce{F}   & $6.68\times10^{-9}$ \\
\enddata
\end{deluxetable}

In order to determine the effects of the reactions listed \response{in Table \ref{tab:reactionTable}} on the Z/E-cyanomethanimine abundance ratio under real interstellar conditions, we have run astrochemical models replicating the conditions of the extended, quiescent part of G+0.693.  For this, we have used the Nautilus v.1.1 astrochemical code \citep{ruaud_gas_2016}, a ``three-phase'' model that simulates reactions in the gas, as well as the ice surface and bulk.

The physical conditions and initial elemental abundances used in our simulations are given in Tables \ref{tab:physparameters} and \ref{tab:initialelemental}, respectively. A gas density of $\sim10^4$ cm$^{-3}$, \responsetwo{based on the work by} \citet{rodriguez-fernandez_non-equilibrium_2000}, was used here. Similarly, we have chosen a value of $T_\mathrm{g}=150$ K, which is at the upper end of the $\sim$50 - 150 K range of kinetic gas temperatures that have been inferred in previous studies \citep{guesten_temperature_1985,huettemeister_kinetic_1993,rodriguez-fernandez_warm_2001,ginsburg_dense_2016,krieger_survey_2017,zeng_complex_2018}. Our choice of $T_\mathrm{d}=15$ K is likewise based on work by \citet{rodriguez-fernandez_iso_2004}, who inferred low dust temperatures $\leq30$ K. Finally, cosmic ray ionization rates, $\zeta$, are typically one of the more unconstrained parameters in astrochemical models. Here, we employ a value of $1.3\times10^{-15}$ s$^{-1}$ following \citet{zeng_complex_2018}, who estimated a value of $\sim 1-10\times10^{-15}$ s$^{-1}$ based on work by \citet{fontani_seeds_2017}. This enhanced value of $\zeta$ relative to the commonly-used $10^{-17}$ s$^{-1}$ is in agreement with enhanced values that have been inferred in other sources near the Galactic Center \citep{ao_thermal_2013,yusef-zadeh_widespread_2013,yusef-zadeh_interacting_2013,shingledecker_inference_2016}. \response{Finally, we should note that the high $A_\mathrm{V}=10$ we use here may seem incompatible with the kinetic temperatures we assume of 150 K; however, previous observational work by \citet{rivilla_abundant_2019} and \citet{zeng_complex_2018} both confirmed high \ce{H2} column densities, and therefore $A_\mathrm{V}$, in G+0.693. Given the location of the source in the Galactic Center, the elevated gas temperature can be understood as arising due to the higher number of shocks \citep{zeng_complex_2018} and enhanced cosmic ray ionization rate characteristic of environments in that region of the Galaxy \citep{ao_thermal_2013}, both of which can be efficient mechanisms for heating in the ISM (see e.g. \citet{burkhardt_modeling_2019,ivlev_gas_2019}).}

\response{For grain processes, we have used the \texttt{NAUTILUS} v1.1 code in its three-phase, i.e. gas/ice surface/ice bulk, mode as previously described by \citet{ruaud_gas_2016}. Specifically, we assume (\textit{i}) a uniform classical grain radius of 0.1 $\mu$m, (\textit{ii}) a surface site density of $1.5\times10^{15}$ sites cm$^{-2}$, (\textit{iii}) a surface comprised of the outer two monolayers following \citet{fayolle_laboratory_2011}, (\textit{iv}) a bulk diffusion mechanism based on the self-movement of water ice, following \citet{ghesquiere_diffusion_2015}, and (\textit{v}) the swapping of material between the surface and bulk using the modified method of \citet{garrod_three-phase_2013} discussed by \citet{ruaud_gas_2016}.}

\begin{figure}[htb]
    \centering
    \plottwo{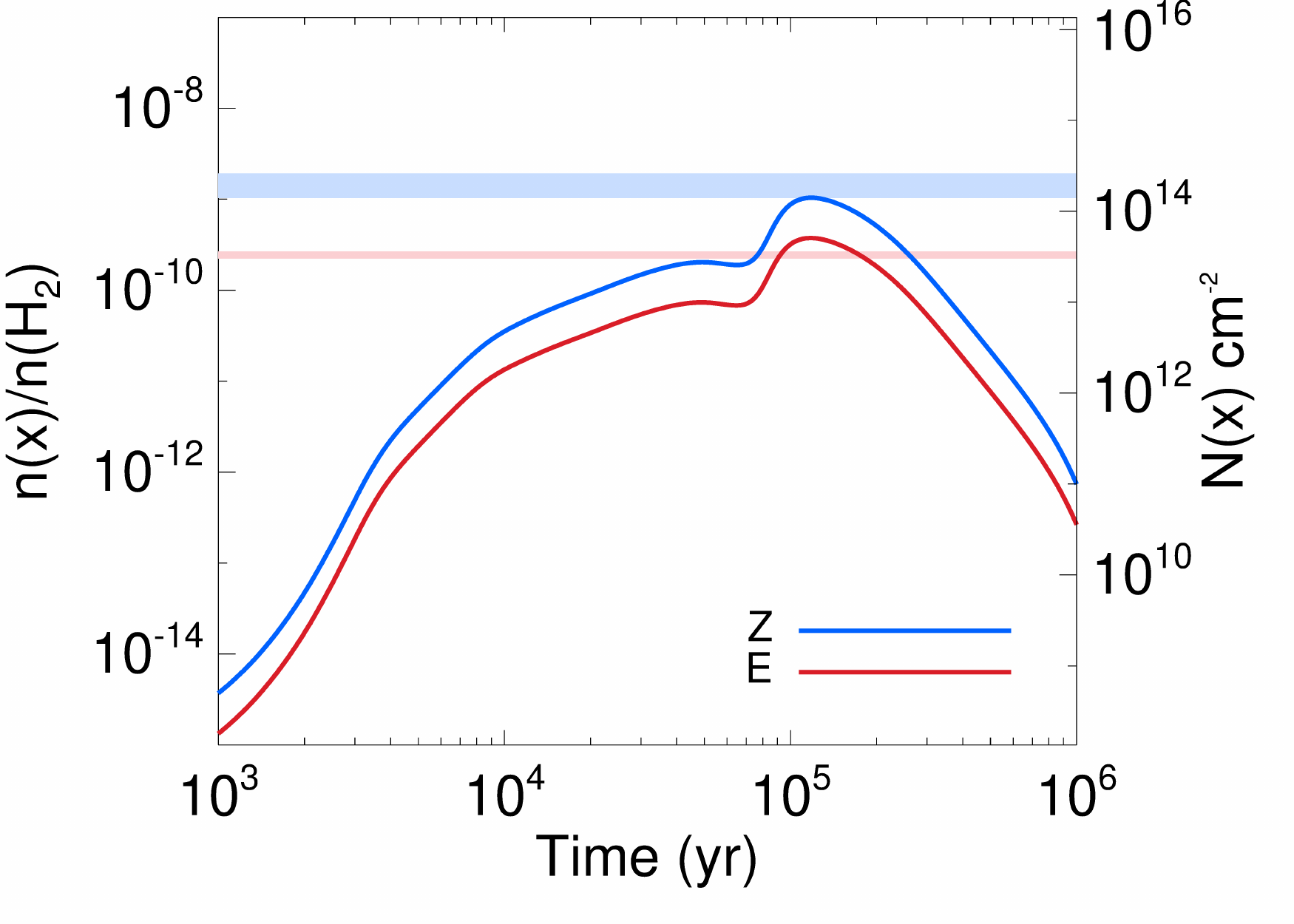}{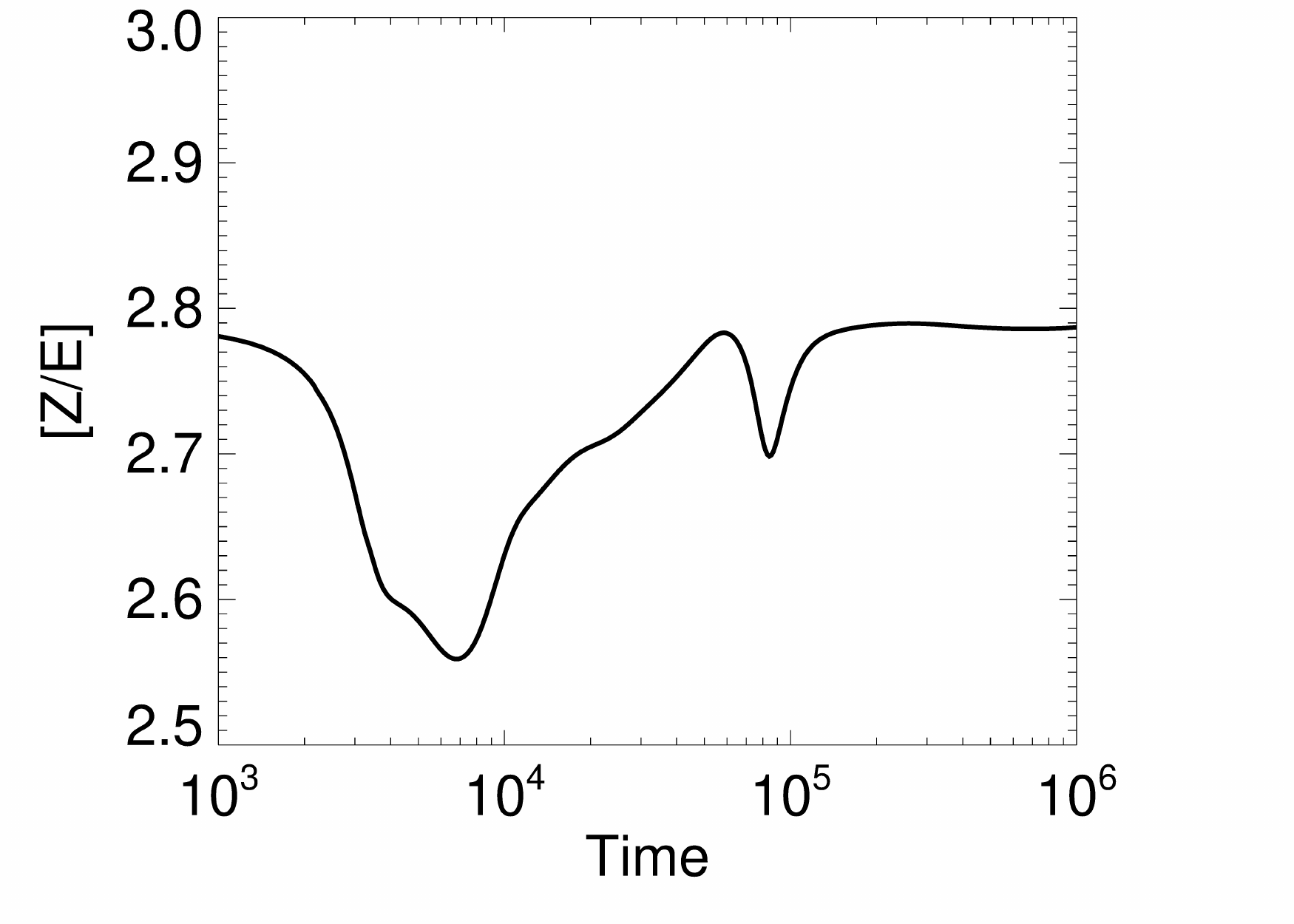}
    \caption{Shown on the left are calculated (solid lines) and observed (horizontal bars) abundances of Z- and E-cyanomethanimine in G+0.693 \firstedit{assuming $T_\mathrm{g}=150$ K, $T_\mathrm{d}=15$ K, and $\zeta=1.3\times10^{-15}$ s$^{-1}$}. Shown on the right is the evolution of [Z/E] during the simulation. Note, error bars for E-cyanomethanimine in the left image are not visible at the scale shown here.}
    \label{fig:modelresults}
\end{figure}

Our chemical network is based on that of \citet{majumdar_methyl_2018} with added reactions from \citet{vastel_isocyanogen_2019} for NCCN/CNCN/\ce{HC2N2+} chemistry. To this base network, we have added reactions relevant to Z/E-cyanomethanimine. \response{In all our models presented here}, Z/E-cyanomethanimine is produced \response{mainly} in the gas via reaction \eqref{R1}, with rates taken from \citet{vazart_cyanomethanimine_2015}, as well as on grains via \eqref{R2} \response{to a lesser extent}, with branching fractions taken from \citet{shivani_formation_2017}. Thus produced, Z- and E-cyanomethanimine are \response{mainly} destroyed via (a) ion-neutral reactions using the total dipoles measured by \citet{takano_microwave_1990} and a dipole polarizability of $\alpha=5.01$ \AA$^3$, estimated using Chemicalize\footnote{\url{https://chemicalize.com}}, (b) photodissociation by internal UV photons using identical order-of-magnitude rates, and (c) the Addition-X and \ce{H2}-abstraction-X reactions described here. 

\response{In adding reactions R3 - R6 to our network, we have assumed that they occur both in the gas and on grain surfaces via a diffusive Langmuir-Hinshelwood mechanism. For reactions on grain surfaces, the rate coefficients are functions of the activation energies, $E_\mathrm{A}$, if present, and the barriers against diffusion, $E_\mathrm{b}$, which we assume are 40\% of the desorption energies, $E_\mathrm{D}$. Here, we have utilized our calculated activation energies listed in Table 2 and 3 for reactions R3 and R5, respectively. Moreover, using desorption energies of $E_\mathrm{D}(\ce{HCN})=3700$ K and $E_\mathrm{D}(\ce{H})=650$ K \citep{wakelam_binding_2017},
we have estimated the desorption energies of $E_\mathrm{D}(\ce{Z/E-HNCHCN})=7400\;\mathrm{K}=2\times E_\mathrm{D}(HCN)$, $E_\mathrm{D}(\ce{H2NCHCN})=8050\;\mathrm{K}=2\times E_\mathrm{D}(HCN)+E_\mathrm{D}(\ce{H})$, and $E_\mathrm{D}(\ce{H3NCHCN})=8050\;\mathrm{K}=2\times E_\mathrm{D}(HCN)+ 2\times E_\mathrm{D}(\ce{H})$. In the case of the H-addition reactions, the dust-grain ice mantle acts as a third body and can stabilize the resulting association products.}

\response{For the gas-phase versions of R3 and R5, we have fit the calculated instanton-corrected bimolecular rate coefficients shown in Figs. 3 and 5 with the modified version of the Arrhenius-Kooij formula originally proposed by \citet{zheng_kinetics_2010} and previously used by us in \citet{shingledecker_efficient_2020}, namely:}

\begin{equation}
  k_\mathrm{inst} = \alpha \left(\frac{T}{300\;\mathrm{K}}\right)^{\beta} \mathrm{exp}\left(-\gamma\frac{T + T_0}{T^2 + T_0^2}\right) \mathrm{cm^{3}\;s^{-1}}.
\label{kinstanton}
\end{equation}

\noindent
\response{Equation \eqref{kinstanton} differs from the standard Arrhenius-Kooij expression in the addition of a $T_0$ term, here assumed to be 150 K, that accounts for the increased efficiency, at low temperatures, of the tunneling-corrected rates. Table \ref{tab:reactionTable} lists the $\alpha$, $\beta$, and $\gamma$ values for R3 and R5. Since, in the gas-phase, the resulting association product of the H-addition reactions to Z/E-cyanomethanimine, \ce{H2NCHCN}, will likely dissociate rather than radiatively stabilize, we assume the ultimate products of R3 in the gas are \ce{NH3 + CCN}. Similarly for reaction R4, we assume the gas-phase products are \ce{H2 + Z/E-HNCHCN}.}

\begin{figure}
\gridline{\fig{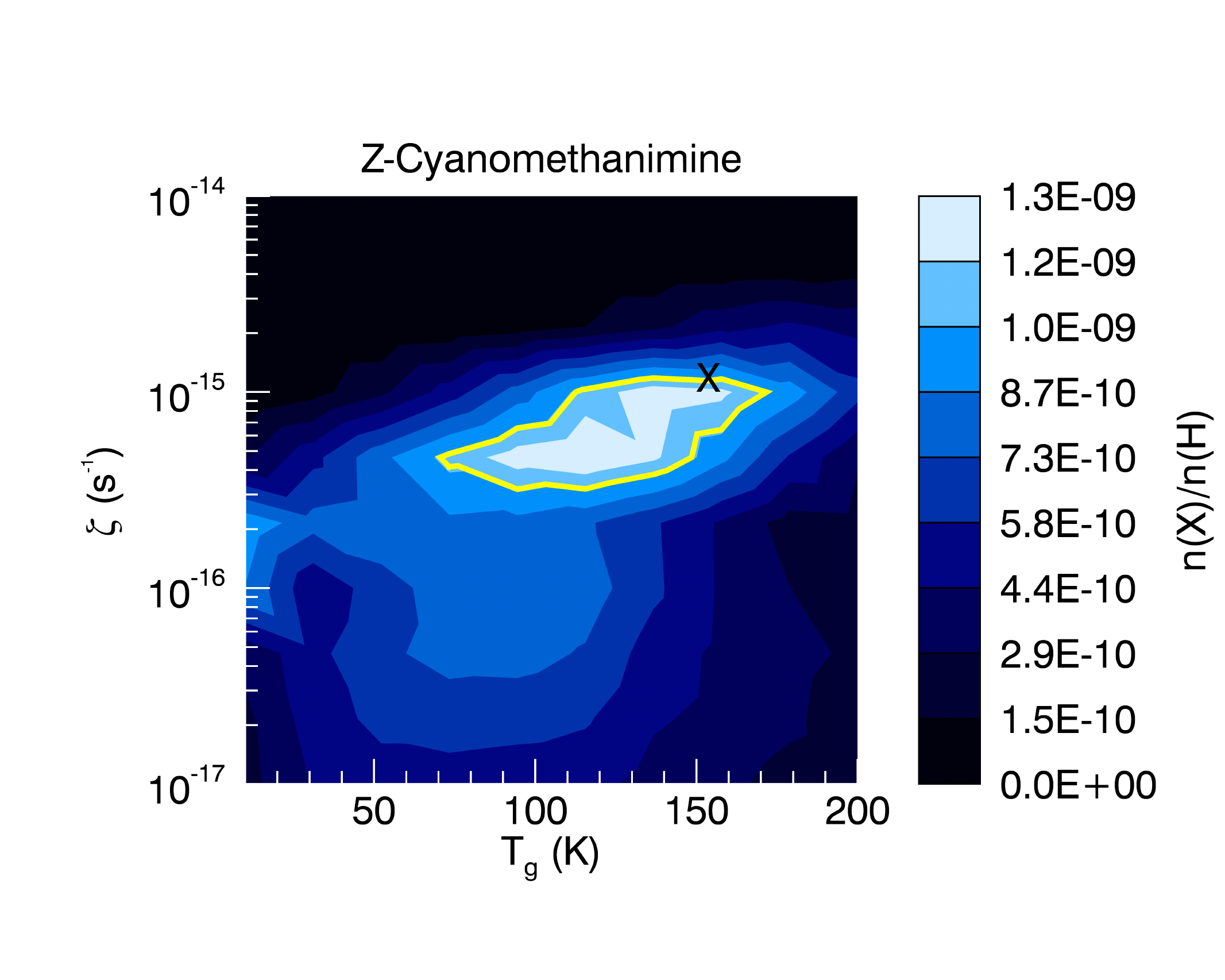}{0.49\textwidth}{(a) \label{fig:zcontour}}
          \fig{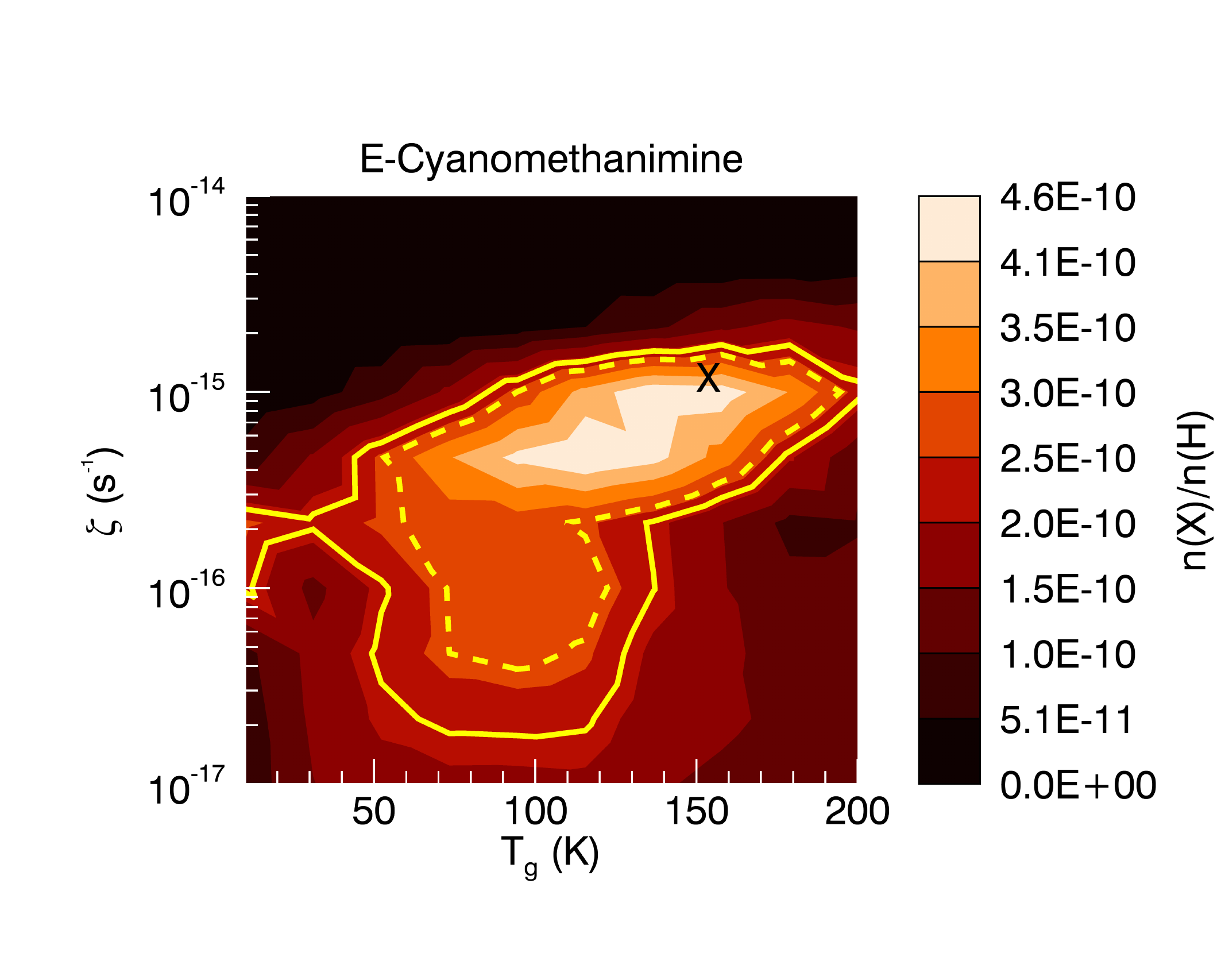}{0.49\textwidth}{(b) \label{fig:econtour}}
          }
\gridline{\fig{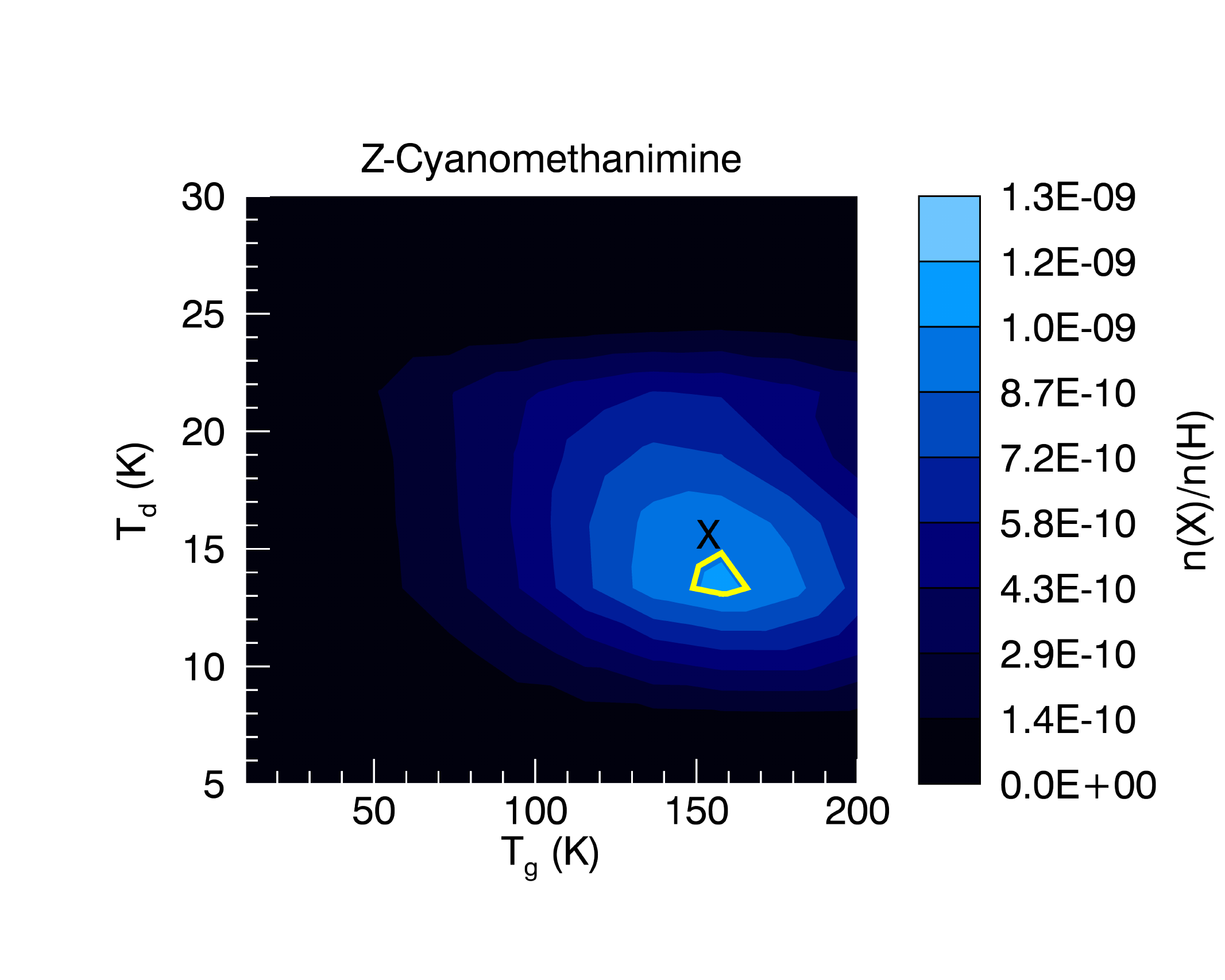}{0.49\textwidth}{(d) \label{fig:zcontour_b}}
          \fig{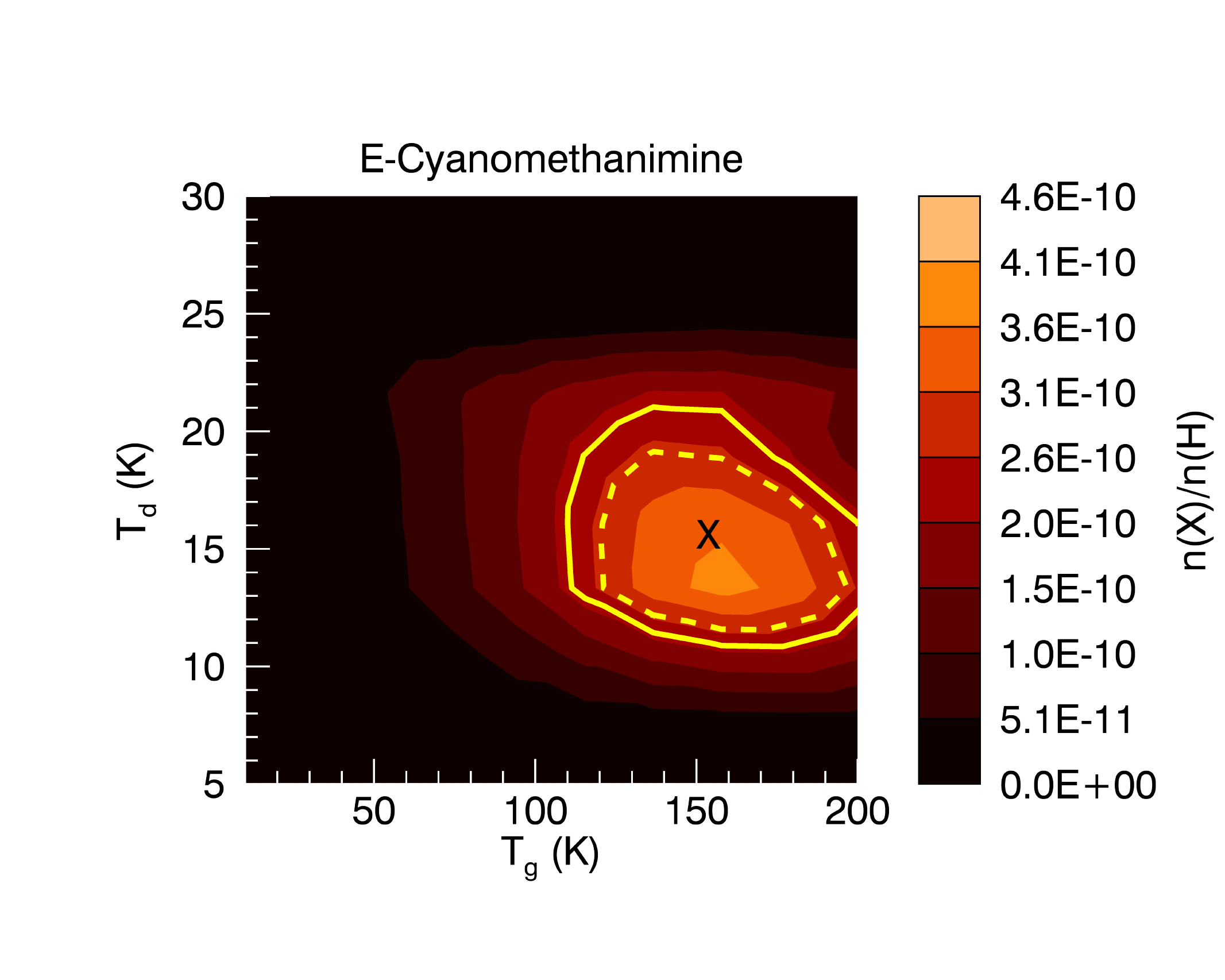}{0.49\textwidth}{(e) \label{fig:econtour_b}}
          }
\caption{Sensitivity of the abundances of Z/E-cyanomethanimine to the cosmic ray ionization rate ($\zeta$), gas temperature ($T_\mathrm{g}$), and dust temperature ($T_\mathrm{d}$). \response{Upper and lower observational limits from \citet{rivilla_abundant_2019} are represented by dashed and solid yellow contours, respectively, and parameters used in our fiducial simulation are indicated by the black ``X''. Note: Figs. (a) and (b) use a constant $T_\mathrm{d}=15$ K, while Figs. (d) and (e) use a constant $\zeta=1.3\times10^{-15}$ s$^{-1}$}.
\label{fig:contours}
}
\end{figure}

The results of our simulations are shown in Fig. \ref{fig:modelresults}. As one can see \firstedit{from the lefthand figure}, the individual abundances of Z- and E-cyanomethanimine nicely match the observational results to within a factor of a few. \firstedit{At all model times, the gas phase reaction \eqref{R1} serves as the main production route for both species. Similarly, throughout our simulation, Z/E-\ce{H2C2N2} are destroyed mainly via reaction with ions. After peaking at around $10^5$ yr, the abundances of both conformers begin to drop as the majority of the carbon, which is at that point mostly in the form of CO, begins to freeze onto grains.} 

As noted though, a fairly wide range of gas temperatures have been inferred for G+0.693, \firstedit{and only upper limits to the dust temperatures were derived by \citet{rodriguez-fernandez_iso_2004}}. Additionally, since a precise method of determining the cosmic ray ionization rate in dense molecular clouds remains elusive \citep{indriolo_cosmic-ray_2013,shingledecker_inference_2016,shingledecker_cosmic-ray-driven_2018}, this key parameter could potentially vary by more than an of magnitude from the value used to obtain Fig. \ref{fig:modelresults}. 

In order to determine the sensitivity of our results to these parameters, we have run a grid of models over a range of gas temperatures ($T_\mathrm{g}\in [10,200]$ K), dust temperatures ($T_\mathrm{d} \in [5,30]$ K), and cosmic ray ionization rates ($\zeta \in [10^{-17},10^{-14}]$ s$^{-1}$), the results of which are shown in Fig. \ref{fig:contours}. \response{There, each point in the grid corresponds to the peak abundance reached during the simulation, which, as illustrated in Fig. \ref{fig:modelresults}, typically corresponded to model times of $\sim10^5$ yr.} \firstedit{Interestingly, there one can see that in all cases there is a well-defined area of the parameter space which yields the highest abundances of Z- and E-cyanomethanimine, and that this region overlaps more-or-less with the previously derived conditions for G+0.693. One can further see that the physical parameters used to obtain the results shown in Fig. \ref{fig:modelresults} correspond roughly with those where the abundances of the two conformers peak.}

\begin{figure}[ht]
\plottwo{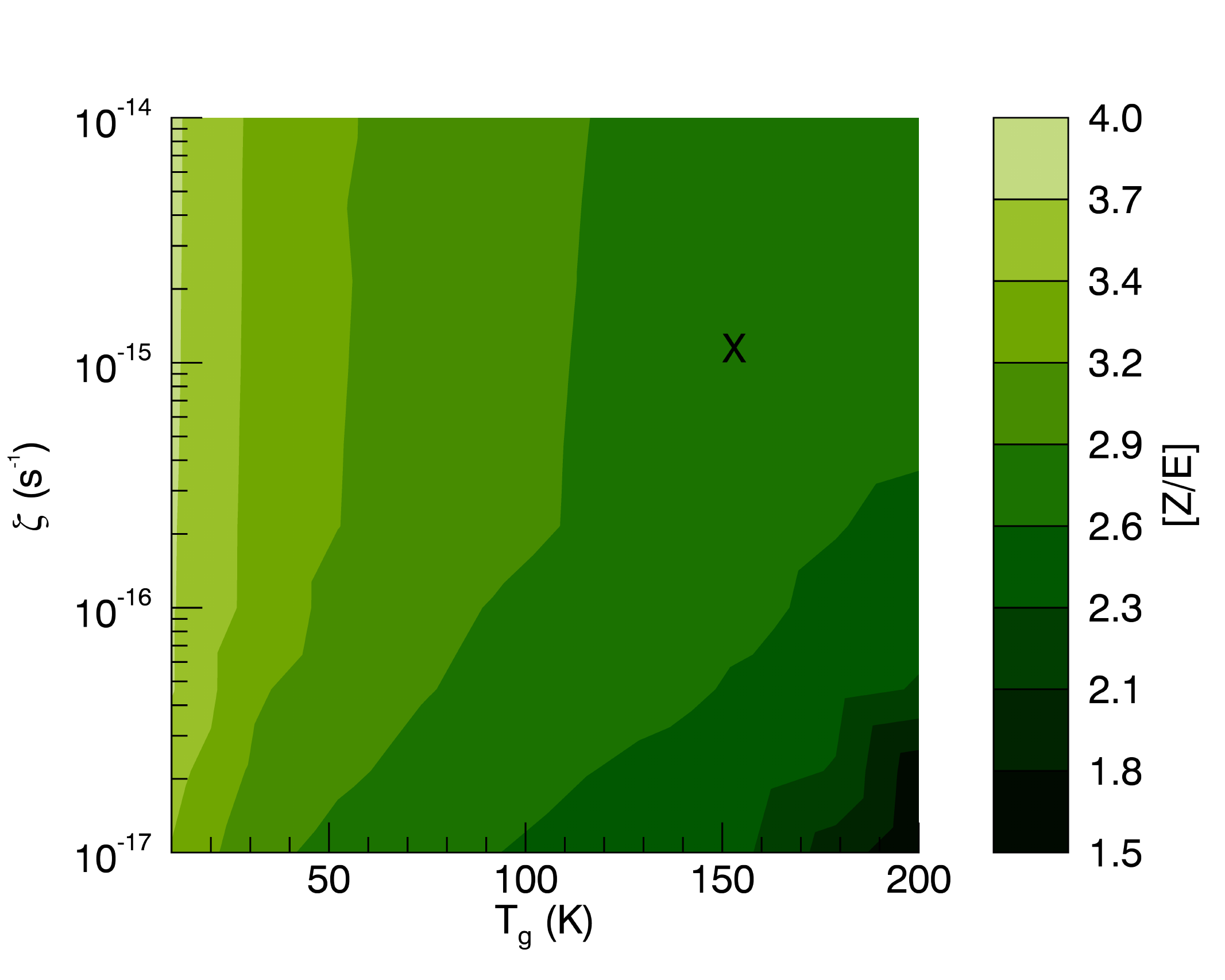}{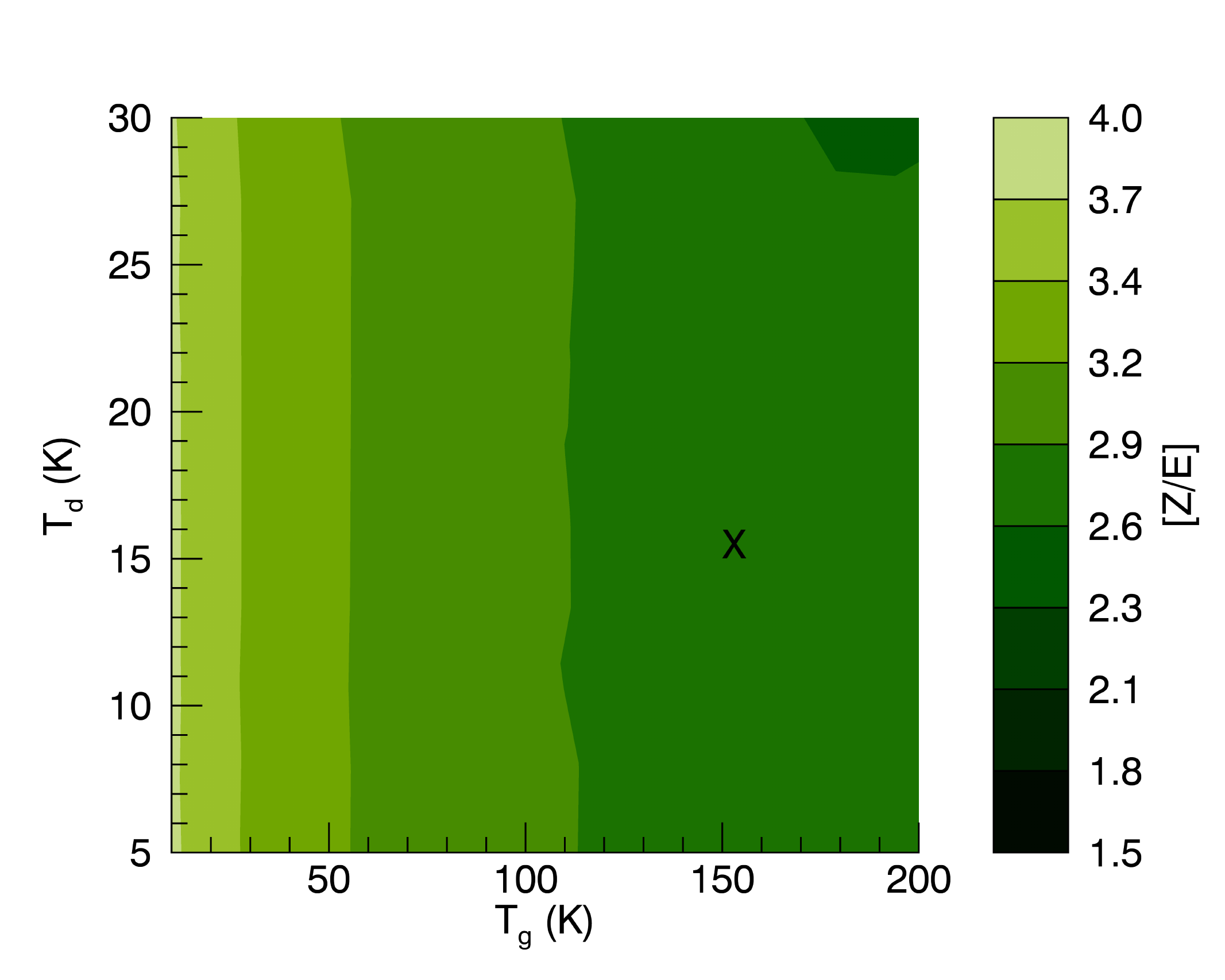}
\caption{Calculated [Z/E] ratios as a function of the gas temperature, $T_\mathrm{g}$ and cosmic ray ionization rate, $\zeta$, (left) and the gas temperature and dust temperature, $T_\mathrm{d}$ (right). \response{Parameters used in our fiducial simulation are indicated by the black ``X''. We note that, here, the dust temperature and cosmic-ray ionisation rates are fixed in the left- and right-hand plots, respectively.}}
\label{fig:ratiocontour}
\end{figure}

\subsection{[Z/E] Ratio}

\firstedit{The resulting [Z/E] values from Fig. \ref{fig:contours} are shown in Fig. \ref{fig:ratiocontour}. There, the left-hand figure shows the abundances of \ref{fig:contours}(a) over \ref{fig:contours}(b), and the right-hand figure those of \ref{fig:contours}(c) over \ref{fig:contours}(d). In both cases, it is obvious that [Z/E] depends most on the gas temperature and least on the grain temperature. The latter finding makes sense, given that the grain-surface formation route \eqref{R2} is more-or-less equally likely to produce either conformer.}

\begin{figure}[htb]
    \plottwo{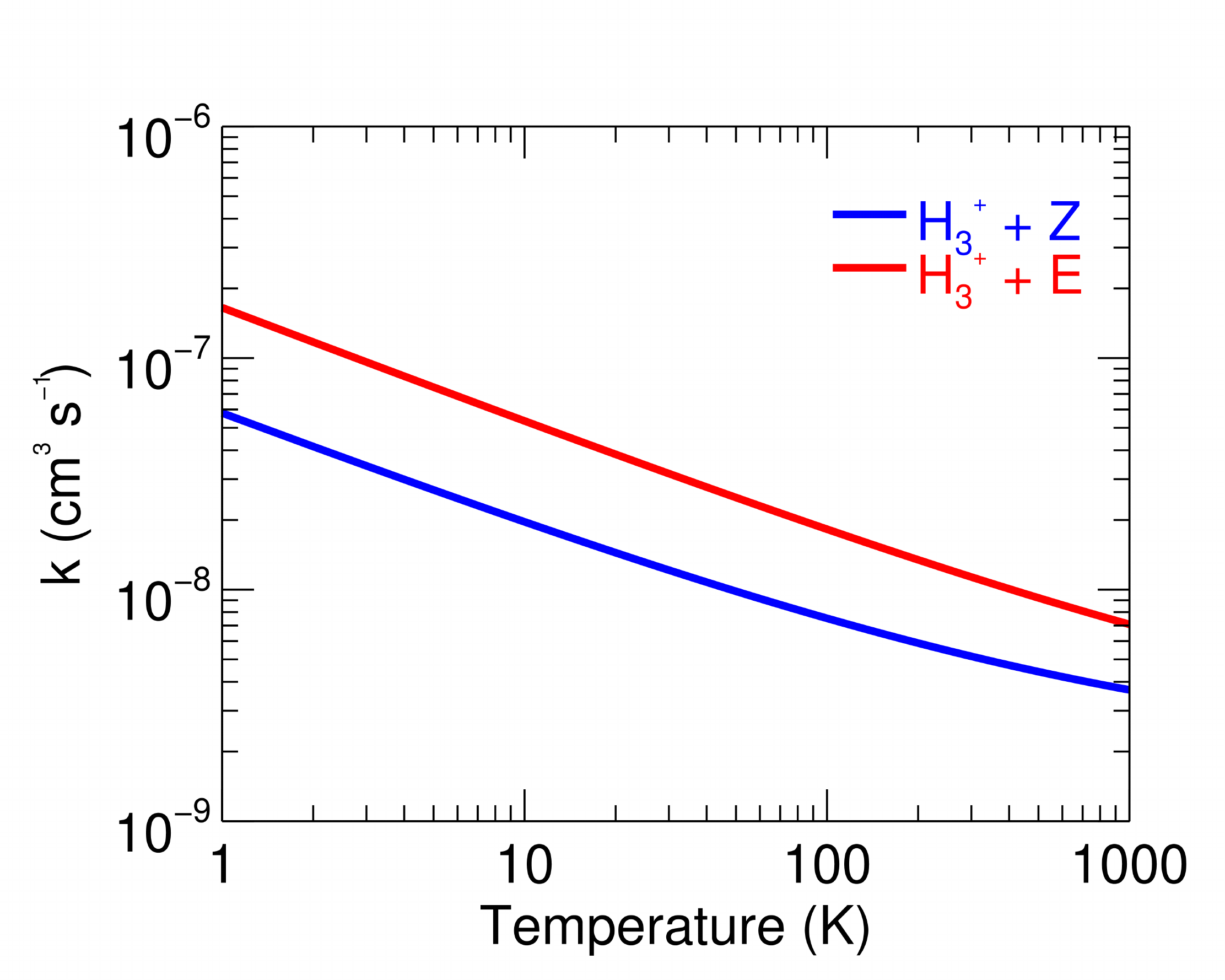}{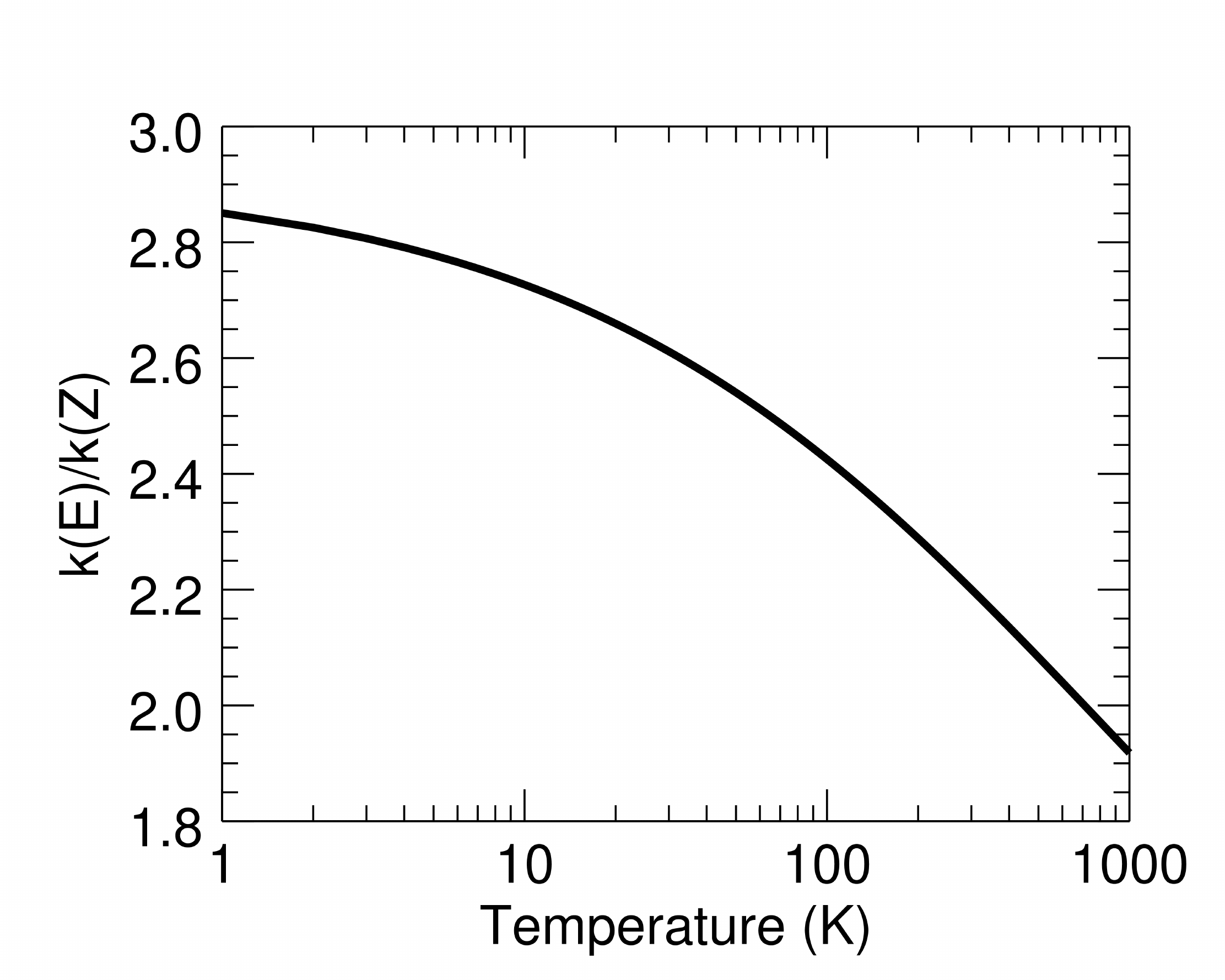}
    \caption{Shown on the left are rate coefficients for the reaction between \ce{H3+ + Z/E-HNCHCN}, calculated using Eq. \eqref{kdlow} as a function of temperature. On the right is the ratio of the rate coefficients for Z- and E-cyanomethanimine.}
    \label{fig:ionrates}
\end{figure}

This strong dependence of the [Z/E] ratios on $T_\mathrm{g}$ is caused by the underlying mechanisms most responsible for the difference in abundance between the two conformers in our model, namely, ion-neutral reactions. As the results of the calculations reported here show, though the two isomers do show somewhat differing reactivity with atomic hydrogen, the overall bimolecular reaction rate coefficients - which are on the order of $\sim10^{-16}$ cm$^3$ s$^{-1}$ - are still much slower than those of fast ion-neutral reactions such as \ce{H3+ + Z/E-HNCHCN}, the rate coefficients of which are shown in Fig. \ref{fig:ionrates}. The difference in rate coefficients observable in Fig. \ref{fig:ionrates} is due to the difference in the permanent dipoles of the two conformers, an effect that is inversely proportional to the gas temperature. The temperature-dependent ion destruction rates are thus what drive the behavior in [Z/E] ratios seen in Fig. \ref{fig:ratiocontour}, and result in values which are quite similar to the ratio of $\mu_\mathrm{E}/\mu_\mathrm{Z} = 4.11/1.41\;\mathrm{D} = 2.9$ as measured by \citet{takano_microwave_1990}. In Fig. \ref{fig:modelresults}, the [Z/E] ratio is 2.8 at the time corresponding to the peak abundance, roughly half of the [Z/E] = \zeratio inferred by \citet{rivilla_abundant_2019}. What is clear is that neither the formation route \eqref{R1} studied by \citet{vazart_cyanomethanimine_2015} ([Z/E]=1.5) or \eqref{R2} studied by \citet{shivani_formation_2017} ([Z/E]$\approx 1$) dominate the [Z/E] ratio in our simulations or the previous observations. 

\response{From Fig. \ref{fig:ratiocontour}, one can see that at low gas temperatures of $\sim10-30$ K, a maximum [Z/E] of $\sim4$ is reached in our models - a value that is within the errors of the [Z/E]$=$\zeratio obtained by \citet{rivilla_abundant_2019}. Further comparison with panels (a) and (b) of Fig. \ref{fig:contours} show that a non-negligible abundance is reached around $\zeta=2\times10^{-16}$ s$^{-1}$ and $T_\mathrm{d}=T_\mathrm{g}=15$ K, where our simulations predict peak abundances of Z/E-cyanomethanimine that are also reasonably close to the observations. However,  gas-temperature measurements of G+0.693 by \citet{zeng_complex_2018} using \ce{CH3CN} - a good tracer of temperature - confirmed previous findings of elevated values of $T_\mathrm{g}$ characteristic of the Galactic Center \citep{guesten_temperature_1985,huettemeister_kinetic_1993,rodriguez-fernandez_warm_2001,ginsburg_dense_2016,krieger_survey_2017}}

\response{Thus, our somewhat lower [Z/E] value at $T_\mathrm{g}=150$ K compared with the observational one indicates some shortcoming on the part of our models. One likely possibility is the absence of important formation/destruction routes in our current chemical network. From Fig. \ref{fig:modelresults}, one can see that agreement between the calculated and observational results could be increased with either the addition of some preferential production pathway for Z-cyanomethanimine, and/or a preferential destruction mechanism for the E isomer. Uncertainties in the model parameters and physical processes simulated in the code also represent another likely source of error in our simulations, particularly those aspects that might be affected by the enhanced cosmic ray ionization rate towards the Galactic Center. For example, the interactions between ice-covered dust grains and cosmic rays can drive a number of physicochemical processes, such as non-thermal chemical reactions \citep{shingledecker_general_2018,shingledecker_cosmic-ray-driven_2018} and the thermal desorption of surface species due to grain heating \citep{bringa_new_2004,ivlev_impulsive_2015}. However, the latter process, in particular, is dependent on grain size \citep{ivlev_gas_2019}, and here we have made the approximation that all grains have the classical radius of 0.1 $\mu$m}

\FloatBarrier
\section{Conclusions} \label{sec:conclusions}

In this work, we report on the results of quantum chemical calculations we have carried on the reaction between atomic hydrogen and Z/E-cyanomethanimine, with the goal of understanding the chemical mechanisms that give rise the [Z/E]$=$\zeratio observed by \citet{rivilla_abundant_2019}. Our main conclusions are the following:

\begin{itemize}
    \item That the reaction of E/Z-cyanomethanimine with H leading to \ce{H2} elimination also proceeds via similar activation energy barriers, but that the resulting bimolecular rate constant for the destruction of E is $\sim1$ order of magnitude larger than for Z. \response{However, the resulting bimolecular rate coefficients for all these reactions was found to be low ($\sim 10^{-17}-10^{-18}$ 
$^3$ s$^{-1}$), and thus will be of comparatively minor importance under astrophysical conditions};
    \item That the Z/E-cyanomethanimine abundance ratio is influenced most strongly by differing ion-polar neutral destruction rates arising from the different permanent dipoles of the two conformers.
\end{itemize}

The last point regarding the relationship between permanent dipole strengths and ion-polar neutral reaction destruction rates can perhaps serve as the basis of a general starting point for either \textit{a priori} predictions of the relative abundances of isomers, or making sense of observational data \textit{ex post facto}. \response{This general ``rule-of-thumb''}, which we shall refer to as the relative dipole principle or RDP, is this: that when the chemistry of a family of isomers is otherwise broadly similar, as appears to be the case with Z/E-cyanomethanimine, then the relative abundances of the different species should be predictable based on - and follow the inverse trend as - the relative magnitudes of their permanent dipole moments. 

Of course, interstellar isomers will not always follow the trend one would expect based on the RDP. As we have shown with propadienone and propynal, two members of the \ce{H2C3O} family of molecules, isomers can display markedly different reactivities with key interstellar species - in this case atomic hydrogen - that can alter the trend one would otherwise predict from the RDP \citep{shingledecker_case_2019}.  However, one advantage of the RDP is that it can yield valuable chemical insights even when it fails, unlike previous attempts at a general rule regarding isomer abundances from thermodynamic arguments such as the minimum energy principle \citep{lattelais_interstellar_2009,lattelais_new_2010}. This is because, due to the enhanced long-range dispersion forces, the underlying ion-polar neutral reactions are quite efficient in interstellar environments and will always be a major part of the chemistry of species with permanent dipoles in the ISM. Therefore, when the RDP fails, it does not mean that ion-polar neutral reactions \textit{aren't} important, rather, it indicates some \textit{even more} important major difference in the chemistry of the isomers, e.g. a missing destruction route or inaccurate branching fractions for key formation routes. \response{It is certainly true that the visibility of interstellar species is dependent on many factors not considered here, such as the properties of non-reactive collisions; nevertheless, since among all factors governing the abundances of isomers in space, the effects summarized by the RDP will usually be a key contributing factor, and since the contribution from ion-neutral destruction reactions are easy to estimate using astrochemical models, as shown here, we feel that our rule-of-thumb provides a useful starting point for a more in-depth investigation.} 

\firstedit{Thus, this work represents the beginning of a series of studies in which we plan to continue our investigations of interstellar isomers. In subsequent works, we will both (a) explore the chemistry of other families of isomers from a theoretical and observational standpoint, as well as (b) continue to explore the validity, limitations, and exceptions of the RDP. Moreover, given the its high sensitivity, future ALMA observations of Z/E-cyanomethanimine and other interstellar isomers in sources with a range of temperatures would also be helpful in determining the role of ion-neutral reactions on their abundances, particularly for families of molecules where the less-stable species also have smaller permanent dipoles.}

\section*{Acknowledgements}

C.N.S. thanks the Alexander von Humboldt foundation for their generous support as well as V. Wakelam for use of the Nautilus v1.1 code.
V.M.R. has received funding from the European Union's Horizon 2020 research and innovation programme under the Marie Sk\l{}odowska-Curie grant agreement No 664931. G.M is also indebted to the Alexander von Humboldt foundation for their support. G.M and J.K would  like  to  acknowledge  funding  from  the  European Union’s Horizon 2020 research and innovation programme (grant agreement No. 646717, TUNNELCHEM) and also the support with computer time by the state of Baden-W\"urttemberg through bwHPC and the German Research Foundation (DFG) through grant no. INST 40/467-1FUGG. 

\software{Nautilus v1.1 \citep{ruaud_gas_2016}, ChemShell \citep{Sherwood2003,Metz2014}, DL-Find \citep{kastner_dl-find:_2009}, Gaussian09 (revision D.01) \citep{g09}, Molpro2015 \cite{Molpro,molpro2015}}.

\bibliographystyle{aasjournal}
\bibliography{merged}

\begin{thebibliography}{}
\expandafter\ifx\csname natexlab\endcsname\relax\def\natexlab#1{#1}\fi
\providecommand{\url}[1]{\href{#1}{#1}}
\providecommand{\dodoi}[1]{doi:~\href{http://doi.org/#1}{\nolinkurl{#1}}}
\providecommand{\doeprint}[1]{\href{http://ascl.net/#1}{\nolinkurl{http://ascl.net/#1}}}
\providecommand{\doarXiv}[1]{\href{https://arxiv.org/abs/#1}{\nolinkurl{https://arxiv.org/abs/#1}}}

\bibitem[{Ao {et~al.}(2013)Ao, Henkel, Menten, Requena-Torres, Stanke,
  Mauersberger, Aalto, Mühle, \& Mangum}]{ao_thermal_2013}
Ao, Y., Henkel, C., Menten, K.~M., {et~al.} 2013, Astronomy and Astrophysics,
  550, A135, \dodoi{10.1051/0004-6361/201220096}

\bibitem[{Belloche {et~al.}(2008)Belloche, Menten, Comito, M{\"{u}}ller,
  Schilke, Ott, Thorwirth, \& Hieret}]{Belloche2008}
Belloche, A., Menten, K.~M., Comito, C., {et~al.} 2008, Astronomy and
  Astrophysics, 482, 179, \dodoi{10.1051/0004-6361:20079203}

\bibitem[{Bringa \& Johnson(2004)}]{bringa_new_2004}
Bringa, E.~M., \& Johnson, R.~E. 2004, The Astrophysical Journal, 603, 159.
\newblock \url{http://iopscience.iop.org/0004-637X/603/1/159}

\bibitem[{Burkhardt {et~al.}(2019)Burkhardt, Shingledecker, Gal, McGuire,
  Remijan, \& Herbst}]{burkhardt_modeling_2019}
Burkhardt, A.~M., Shingledecker, C.~N., Gal, R.~L., {et~al.} 2019, The
  Astrophysical Journal, 881, 32, \dodoi{10.3847/1538-4357/ab2be8}

\bibitem[{Callahan {et~al.}(2011)Callahan, Smith, Cleaves, Ruzicka, Stern,
  Glavin, House, \& Dworkin}]{callahan_carbonaceous_2011}
Callahan, M.~P., Smith, K.~E., Cleaves, H.~J., {et~al.} 2011, Proceedings of
  the National Academy of Sciences, 108, 13995, \dodoi{10.1073/pnas.1106493108}

\bibitem[{Chakrabarti \& Chakrabarti(2000)}]{chakrabarti_can_2000}
Chakrabarti, S., \& Chakrabarti, S.~K. 2000, Astronomy and Astrophysics, 354,
  L6.
\newblock \url{http://adsabs.harvard.edu/abs/2000A%26A...354L...6C}

\bibitem[{Clemmons {et~al.}(1983)Clemmons, Jasien, \&
  Dykstra}]{clemmons_possible_1983}
Clemmons, J.~H., Jasien, P.~G., \& Dykstra, C.~E. 1983, Molecular Physics, 48,
  631, \dodoi{10.1080/00268978300100461}

\bibitem[{Danger {et~al.}(2011)Danger, Bossa, {De Marcellus}, Borget, Duvernay,
  Theul{\'{e}}, Chiavassa, \& D'Hendecourt}]{Danger2011}
Danger, G., Bossa, J.~B., {De Marcellus}, P., {et~al.} 2011, Astronomy and
  Astrophysics, 525, A30, \dodoi{10.1051/0004-6361/201015736}

\bibitem[{Ehrenfreund {et~al.}(2001)Ehrenfreund, Glavin, Botta, Cooper, \&
  Bada}]{ehrenfreund_extraterrestrial_2001}
Ehrenfreund, P., Glavin, D.~P., Botta, O., Cooper, G., \& Bada, J.~L. 2001,
  Proceedings of the National Academy of Sciences, 98, 2138,
  \dodoi{10.1073/pnas.051502898}

\bibitem[{Fayolle {et~al.}(2011)Fayolle, Öberg, Cuppen, Visser, \&
  Linnartz}]{fayolle_laboratory_2011}
Fayolle, E.~C., Öberg, K.~I., Cuppen, H.~M., Visser, R., \& Linnartz, H. 2011,
  Astronomy and Astrophysics, 529, A74, \dodoi{10.1051/0004-6361/201016121}

\bibitem[{Fontani {et~al.}(2017)Fontani, Ceccarelli, Favre, Caselli, Neri,
  Sims, Kahane, Alves, Balucani, Bianchi, Caux, Jaber Al-Edhari,
  Lopez-Sepulcre, Pineda, Bachiller, Bizzocchi, Bottinelli, Chacon-Tanarro,
  Choudhury, Codella, Coutens, Dulieu, Feng, Rimola, Hily-Blant, Holdship,
  Jimenez-Serra, Laas, Lefloch, Oya, Podio, Pon, Punanova, Quenard, Sakai,
  Spezzano, Taquet, Testi, Theulé, Ugliengo, Vastel, Vasyunin, Viti, Yamamoto,
  \& Wiesenfeld}]{fontani_seeds_2017}
Fontani, F., Ceccarelli, C., Favre, C., {et~al.} 2017, Astronomy and
  Astrophysics, 605, A57, \dodoi{10.1051/0004-6361/201730527}

\bibitem[{Frisch {et~al.}(2016)Frisch, Trucks, Schlegel, Scuseria, Robb,
  Cheeseman, Scalmani, Barone, Mennucci, Petersson, Nakatsuji, Caricato, Li,
  Hratchian, Izmaylov, Bloino, Zheng, Sonnenberg, Hada, Ehara, Toyota, Fukuda,
  Hasegawa, Ishida, Nakajima, Honda, Kitao, Nakai, Vreven, Montgomery, Peralta,
  Ogliaro, Bearpark, Heyd, Brothers, Kudin, Staroverov, Kobayashi, Normand,
  Raghavachari, Rendell, Burant, Iyengar, Tomasi, Cossi, Rega, Millam, Klene,
  Knox, Cross, Bakken, Adamo, Jaramillo, Gomperts, Stratmann, Yazyev, Austin,
  Cammi, Pomelli, Ochterski, Martin, Morokuma, Zakrzewski, Voth, Salvador,
  Dannenberg, Dapprich, Daniels, Farkas, Foresman, Ortiz, Cioslowski, \&
  Fox}]{g09}
Frisch, M.~J., Trucks, G.~W., Schlegel, H.~B., {et~al.} 2016, Gaussian09
  {R}evision {D}.01

\bibitem[{Garrod(2013)}]{garrod_three-phase_2013}
Garrod, R.~T. 2013, The Astrophysical Journal, 765, 60,
  \dodoi{10.1088/0004-637X/765/1/60}

\bibitem[{Ghesquière {et~al.}(2015)Ghesquière, Mineva, Talbi, Theulé, Noble,
  \& Chiavassa}]{ghesquiere_diffusion_2015}
Ghesquière, P., Mineva, T., Talbi, D., {et~al.} 2015, Physical Chemistry
  Chemical Physics, 17, 11455, \dodoi{10.1039/C5CP00558B}

\bibitem[{Ginsburg {et~al.}(2016)Ginsburg, Henkel, Ao, Riquelme, Kauffmann,
  Pillai, Mills, Requena-Torres, Immer, Testi, Ott, Bally, Battersby, Darling,
  Aalto, Stanke, Kendrew, Kruijssen, Longmore, Dale, Guesten, \&
  Menten}]{ginsburg_dense_2016}
Ginsburg, A., Henkel, C., Ao, Y., {et~al.} 2016, Astronomy \& Astrophysics,
  586, A50, \dodoi{10.1051/0004-6361/201526100}

\bibitem[{Guesten {et~al.}(1985)Guesten, Walmsley, Ungerechts, \&
  Churchwell}]{guesten_temperature_1985}
Guesten, R., Walmsley, C.~M., Ungerechts, H., \& Churchwell, E. 1985, Astronomy
  and Astrophysics, 142, 381.
\newblock \url{http://adsabs.harvard.edu/abs/1985A%26A...142..381G}

\bibitem[{Hacar {et~al.}(2020)Hacar, Bosman, \& Dishoeck}]{hacar_hcn--hnc_2020}
Hacar, A., Bosman, A.~D., \& Dishoeck, E. F.~v. 2020, Astronomy \&
  Astrophysics, 635, A4, \dodoi{10.1051/0004-6361/201936516}

\bibitem[{Herbst(2006)}]{herbst_gas_2006}
Herbst, E. 2006, in Springer {Handbook} of {Atomic}, {Molecular}, and {Optical}
  {Physics}, ed. G.~Drake, Springer {Handbooks} (New York, NY: Springer),
  561--574, \dodoi{10.1007/978-0-387-26308-3_37}

\bibitem[{Herbst {et~al.}(2000)Herbst, Terzieva, \&
  Talbi}]{herbst_calculations_2000}
Herbst, E., Terzieva, R., \& Talbi, D. 2000, Monthly Notices of the Royal
  Astronomical Society, 311, 869, \dodoi{10.1046/j.1365-8711.2000.03103.x}

\bibitem[{Huettemeister {et~al.}(1993)Huettemeister, Wilson, Bania, \&
  Martin-Pintado}]{huettemeister_kinetic_1993}
Huettemeister, S., Wilson, T.~L., Bania, T.~M., \& Martin-Pintado, J. 1993,
  Astronomy and Astrophysics, 280, 255.
\newblock \url{http://adsabs.harvard.edu/abs/1993A%26A...280..255H}

\bibitem[{Indriolo \& McCall(2013)}]{indriolo_cosmic-ray_2013}
Indriolo, N., \& McCall, B.~J. 2013, Chemical Society Reviews, 42, 7763,
  \dodoi{10.1039/C3CS60087D}

\bibitem[{Irvine \& Schloerb(1984)}]{irvine_cyanide_1984}
Irvine, W.~M., \& Schloerb, F.~P. 1984, The Astrophysical Journal, 282, 516,
  \dodoi{10.1086/162229}

\bibitem[{Ivlev {et~al.}(2015)Ivlev, Röcker, Vasyunin, \&
  Caselli}]{ivlev_impulsive_2015}
Ivlev, A.~V., Röcker, T.~B., Vasyunin, A., \& Caselli, P. 2015, The
  Astrophysical Journal, 805, 59, \dodoi{10.1088/0004-637X/805/1/59}

\bibitem[{Ivlev {et~al.}(2019)Ivlev, Silsbee, Sipilä, \&
  Caselli}]{ivlev_gas_2019}
Ivlev, A.~V., Silsbee, K., Sipilä, O., \& Caselli, P. 2019, The Astrophysical
  Journal, 884, 176, \dodoi{10.3847/1538-4357/ab4252}

\bibitem[{Karton \& Talbi(2014)}]{karton_pinning_2014}
Karton, A., \& Talbi, D. 2014, Chemical Physics, 436, 22,
  \dodoi{10.1016/j.chemphys.2014.03.010}

\bibitem[{K{\"{a}}stner {et~al.}(2009)K{\"{a}}stner, Carr, Keal, Thiel, Wander,
  \& Sherwood}]{Kastner2009}
K{\"{a}}stner, J., Carr, J.~M., Keal, T.~W., {et~al.} 2009, Journal of Physical
  Chemistry A, 113, 11856, \dodoi{10.1021/jp9028968}

\bibitem[{Koch {et~al.}(2008)Koch, Toubin, Peslherbe, \& Hynes}]{Koch2008}
Koch, D.~M., Toubin, C., Peslherbe, G.~H., \& Hynes, J.~T. 2008, The Journal of
  Physical Chemistry C, 112, 2972, \dodoi{10.1021/jp076221+}

\bibitem[{Kolesnikov{\'{a}} {et~al.}(2017)Kolesnikov{\'{a}}, Alonso, Mata, \&
  Alonso}]{Kolesnikova2017}
Kolesnikov{\'{a}}, L., Alonso, E.~R., Mata, S., \& Alonso, J.~L. 2017, The
  Astrophysical Journal Supplement Series, 229, 26,
  \dodoi{10.3847/1538-4365/aa5d13}

\bibitem[{Krieger {et~al.}(2017)Krieger, Ott, Beuther, Walter, Kruijssen,
  Meier, Mills, Contreras, Edwards, Ginsburg, Henkel, Henshaw, Jackson,
  Kauffmann, Longmore, Martín, Morris, Pillai, Rickert, Rosolowsky, Shinnaga,
  Walsh, Yusef-Zadeh, \& Zhang}]{krieger_survey_2017}
Krieger, N., Ott, J., Beuther, H., {et~al.} 2017, The Astrophysical Journal,
  850, 77, \dodoi{10.3847/1538-4357/aa951c}

\bibitem[{Kästner {et~al.}(2009)Kästner, Carr, Keal, Thiel, Wander, \&
  Sherwood}]{kastner_dl-find:_2009}
Kästner, J., Carr, J.~M., Keal, T.~W., {et~al.} 2009, The Journal of Physical
  Chemistry A, 113, 11856, \dodoi{10.1021/jp9028968}

\bibitem[{Lattelais {et~al.}(2009)Lattelais, Pauzat, Ellinger, \&
  Ceccarelli}]{lattelais_interstellar_2009}
Lattelais, M., Pauzat, F., Ellinger, Y., \& Ceccarelli, C. 2009, The
  Astrophysical Journal Letters, 696, L133,
  \dodoi{10.1088/0004-637X/696/2/L133}

\bibitem[{Lattelais {et~al.}(2010)Lattelais, Pauzat, Ellinger, \&
  Ceccarelli}]{lattelais_new_2010}
---. 2010, Astronomy and Astrophysics, 519, A30,
  \dodoi{10.1051/0004-6361/200913869}

\bibitem[{Loison {et~al.}(2016)Loison, Agúndez, Marcelino, Wakelam, Hickson,
  Cernicharo, Gerin, Roueff, \& Guélin}]{loison_interstellar_2016}
Loison, J.-C., Agúndez, M., Marcelino, N., {et~al.} 2016, Monthly Notices of
  the Royal Astronomical Society, 456, 4101, \dodoi{10.1093/mnras/stv2866}

\bibitem[{Loomis {et~al.}(2015)Loomis, McGuire, Shingledecker, Johnson, Blair,
  {Amy Robertson}, \& Remijan}]{loomis_investigating_2015}
Loomis, R.~A., McGuire, B.~A., Shingledecker, C., {et~al.} 2015, The
  Astrophysical Journal, 799, 34.
\newblock \url{http://stacks.iop.org/0004-637X/799/i=1/a=34}

\bibitem[{Majumdar {et~al.}(2018)Majumdar, Loison, Ruaud, Gratier, Wakelam, \&
  Coutens}]{majumdar_methyl_2018}
Majumdar, L., Loison, J.-C., Ruaud, M., {et~al.} 2018, Monthly Notices of the
  Royal Astronomical Society, 473, L59, \dodoi{10.1093/mnrasl/slx157}

\bibitem[{McConnell \& K{\"{a}}stner(2017)}]{McConnell2017}
McConnell, S., \& K{\"{a}}stner, J. 2017, Journal of Computational Chemistry,
  38, 2570, \dodoi{10.1002/jcc.24914}

\bibitem[{Metz {et~al.}(2014)Metz, K{\"{a}}stner, Sokol, Keal, \&
  Sherwood}]{Metz2014}
Metz, S., K{\"{a}}stner, J., Sokol, A.~A., Keal, T.~W., \& Sherwood, P. 2014,
  Wiley Interdisciplinary Reviews: Computational Molecular Science, 4, 101,
  \dodoi{10.1002/wcms.1163}

\bibitem[{Oró(1961)}]{oro_mechanism_1961}
Oró, J. 1961, Nature, 191, 1193, \dodoi{10.1038/1911193a0}

\bibitem[{Padovani {et~al.}(2018)Padovani, Galli, Ivlev, Caselli, \&
  Ferrara}]{padovani_production_2018}
Padovani, M., Galli, D., Ivlev, A.~V., Caselli, P., \& Ferrara, A. 2018,
  Astronomy and Astrophysics, 619, A144, \dodoi{10.1051/0004-6361/201834008}

\bibitem[{Puzzarini(2015)}]{Puzzarini2015}
Puzzarini, C. 2015, Journal of Physical Chemistry A, 119, 11614,
  \dodoi{10.1021/acs.jpca.5b09489}

\bibitem[{Rivilla {et~al.}(2019)Rivilla, Martín-Pintado, Jiménez-Serra, Zeng,
  Martín, Armijos-Abendaño, Requena-Torres, Aladro, \&
  Riquelme}]{rivilla_abundant_2019}
Rivilla, V.~M., Martín-Pintado, J., Jiménez-Serra, I., {et~al.} 2019, Monthly
  Notices of the Royal Astronomical Society: Letters, 483, L114,
  \dodoi{10.1093/mnrasl/sly228}

\bibitem[{Rodríguez-Fernández {et~al.}(2000)Rodríguez-Fernández,
  Martín-Pintado, de~Vicente, Fuente, Hüttemeister, Wilson, \&
  Kunze}]{rodriguez-fernandez_non-equilibrium_2000}
Rodríguez-Fernández, N.~J., Martín-Pintado, J., de~Vicente, P., {et~al.}
  2000, A\&A, 356, 695.
\newblock \url{https://ui.adsabs.harvard.edu/abs/2000A&A...356..695R/abstract}

\bibitem[{Rodríguez-Fernández {et~al.}(2001)Rodríguez-Fernández,
  Martín-Pintado, Fuente, Vicente, Wilson, \&
  Hüttemeister}]{rodriguez-fernandez_warm_2001}
Rodríguez-Fernández, N.~J., Martín-Pintado, J., Fuente, A., {et~al.} 2001,
  Astronomy \& Astrophysics, 365, 174, \dodoi{10.1051/0004-6361:20000020}

\bibitem[{Rodríguez-Fernández {et~al.}(2004)Rodríguez-Fernández,
  Martín-Pintado, Fuente, \& Wilson}]{rodriguez-fernandez_iso_2004}
Rodríguez-Fernández, N.~J., Martín-Pintado, J., Fuente, A., \& Wilson, T.~L.
  2004, Astronomy \& Astrophysics, 427, 217, \dodoi{10.1051/0004-6361:20041370}

\bibitem[{Rommel {et~al.}(2011)Rommel, Goumans, \& K\"astner}]{Rommel2011-2}
Rommel, J.~B., Goumans, T.~P., \& K\"astner, J. 2011, Journal of Chemical
  Theory and Computation, 7, 690, \dodoi{10.1021/ct100658y}

\bibitem[{Rommel \& K{\"{a}}stner(2011)}]{Rommel2011}
Rommel, J.~B., \& K{\"{a}}stner, J. 2011, Journal of Chemical Physics, 134,
  184107, \dodoi{10.1063/1.3587240}

\bibitem[{Ruaud {et~al.}(2016)Ruaud, Wakelam, \& Hersant}]{ruaud_gas_2016}
Ruaud, M., Wakelam, V., \& Hersant, F. 2016, Monthly Notices of the Royal
  Astronomical Society, 459, 3756, \dodoi{10.1093/mnras/stw887}

\bibitem[{Schilke {et~al.}(1992)Schilke, Walmsley, Pineau Des~Forets, Roueff,
  Flower, \& Guilloteau}]{schilke_study_1992}
Schilke, P., Walmsley, C.~M., Pineau Des~Forets, G., {et~al.} 1992, A\&A, 256,
  595.
\newblock \url{https://ui.adsabs.harvard.edu/abs/1992A&A...256..595S/abstract}

\bibitem[{Sherwood {et~al.}(2003)Sherwood, {De Vries}, Guest, Schreckenbach,
  Catlow, French, Sokol, Bromley, Thiel, Turner, Billeter, Terstegen, Thiel,
  Kendrick, Rogers, Casci, Watson, King, Karlsen, Sj{\o}voll, Fahmi,
  Sch{\"{a}}fer, \& Lennartz}]{Sherwood2003}
Sherwood, P., {De Vries}, A.~H., Guest, M.~F., {et~al.} 2003, Journal of
  Molecular Structure: THEOCHEM, 632, 1, \dodoi{10.1016/s0166-1280(03)00285-9}

\bibitem[{Shingledecker {et~al.}(2016)Shingledecker, Bergner, Le~Gal, Öberg,
  Hincelin, \& Herbst}]{shingledecker_inference_2016}
Shingledecker, C.~N., Bergner, J.~B., Le~Gal, R., {et~al.} 2016, The
  Astrophysical Journal, 830, 151, \dodoi{10.3847/0004-637X/830/2/151}

\bibitem[{Shingledecker \& Herbst(2018)}]{shingledecker_general_2018}
Shingledecker, C.~N., \& Herbst, E. 2018, Physical Chemistry Chemical Physics,
  20, 5359, \dodoi{10.1039/C7CP05901A}

\bibitem[{Shingledecker {et~al.}(2020)Shingledecker, Lamberts, Laas, Vasyunin,
  Herbst, Kästner, \& Caselli}]{shingledecker_efficient_2020}
Shingledecker, C.~N., Lamberts, T., Laas, J.~C., {et~al.} 2020, The
  Astrophysical Journal, 888, 52, \dodoi{10.3847/1538-4357/ab5360}

\bibitem[{Shingledecker {et~al.}(2019)Shingledecker, Álvarez Barcia, Korn, \&
  Kästner}]{shingledecker_case_2019}
Shingledecker, C.~N., Álvarez Barcia, S., Korn, V.~H., \& Kästner, J. 2019,
  The Astrophysical Journal, 878, 80, \dodoi{10.3847/1538-4357/ab1d4a}

\bibitem[{Shingledecker {et~al.}(2018)Shingledecker, Tennis, Gal, \&
  Herbst}]{shingledecker_cosmic-ray-driven_2018}
Shingledecker, C.~N., Tennis, J., Gal, R.~L., \& Herbst, E. 2018, The
  Astrophysical Journal, 861, 20, \dodoi{10.3847/1538-4357/aac5ee}

\bibitem[{{Shivani} {et~al.}(2017){Shivani}, Misra, \&
  Tandon}]{shivani_formation_2017}
{Shivani}, Misra, A., \& Tandon, P. 2017, Research in Astronomy and
  Astrophysics, 17, 1, \dodoi{10.1088/1674-4527/17/1/1}

\bibitem[{Su \& Chesnavich(1982)}]{su_parametrization_1982}
Su, T., \& Chesnavich, W.~J. 1982, Journal of Chemical Physics, 76, 5183,
  \dodoi{10.1063/1.442828}

\bibitem[{Takano {et~al.}(1990)Takano, Sugie, Sugawara, Takeo, Matsumura,
  Masuda, \& Kuchitsu}]{takano_microwave_1990}
Takano, S., Sugie, M., Sugawara, K.-i., {et~al.} 1990, Journal of Molecular
  Spectroscopy, 141, 13, \dodoi{10.1016/0022-2852(90)90273-S}

\bibitem[{Ungerechts {et~al.}(1997)Ungerechts, Bergin, Goldsmith, Irvine,
  Schloerb, \& Snell}]{ungerechts_chemical_1997}
Ungerechts, H., Bergin, E.~A., Goldsmith, P.~F., {et~al.} 1997, The
  Astrophysical Journal, 482, 245, \dodoi{10.1086/304110}

\bibitem[{Vastel {et~al.}(2019)Vastel, Loison, Wakelam, \&
  Lefloch}]{vastel_isocyanogen_2019}
Vastel, C., Loison, J.~C., Wakelam, V., \& Lefloch, B. 2019, Astronomy \&
  Astrophysics, 625, A91, \dodoi{10.1051/0004-6361/201935010}

\bibitem[{Vazart {et~al.}(2015)Vazart, Latouche, Skouteris, Balucani, \&
  Barone}]{vazart_cyanomethanimine_2015}
Vazart, F., Latouche, C., Skouteris, D., Balucani, N., \& Barone, V. 2015, The
  Astrophysical Journal, 810, 111, \dodoi{10.1088/0004-637X/810/2/111}

\bibitem[{Wakelam {et~al.}(2017)Wakelam, Loison, Mereau, \&
  Ruaud}]{wakelam_binding_2017}
Wakelam, V., Loison, J.~C., Mereau, R., \& Ruaud, M. 2017, Molecular
  Astrophysics, 6, 22, \dodoi{10.1016/j.molap.2017.01.002}

\bibitem[{Weigend \& Ahlrichs(2005)}]{Weigend2005}
Weigend, F., \& Ahlrichs, R. 2005, Physical Chemistry Chemical Physics, 7,
  3297, \dodoi{10.1039/B508541A}

\bibitem[{Werner {et~al.}(2012)Werner, Knowles, Knizia, Manby, \&
  Sch{\"{u}}tz}]{Molpro}
Werner, H.-J., Knowles, P.~J., Knizia, G., Manby, F.~R., \& Sch{\"{u}}tz, M.
  2012, WIREs Comput Mol Sci, 2, 242

\bibitem[{Werner {et~al.}(2015)Werner, Knowles, Knizia, Manby, {Sch\"{u}tz},
  Celani, Gy\"orffy, Kats, Korona, Lindh, Mitrushenkov, Rauhut, Shamasundar,
  Adler, Amos, Bernhardsson, Berning, Cooper, Deegan, Dobbyn, Eckert, Goll,
  Hampel, Hesselmann, Hetzer, Hrenar, Jansen, K\"oppl, Liu, Lloyd, Mata, May,
  McNicholas, Meyer, Mura, Nicklass, O'Neill, Palmieri, Peng, Pfl\"uger,
  Pitzer, Reiher, Shiozaki, Stoll, Stone, Tarroni, Thorsteinsson, \&
  Wang}]{molpro2015}
Werner, H.-J., Knowles, P.~J., Knizia, G., {et~al.} 2015, MOLPRO, version
  2015.1, a package of ab initio programs

\bibitem[{Woon \& Herbst(2009)}]{woon_quantum_2009}
Woon, D.~E., \& Herbst, E. 2009, The Astrophysical Journal Supplement Series,
  185, 273, \dodoi{10.1088/0067-0049/185/2/273}

\bibitem[{Yusef-Zadeh {et~al.}(2013{\natexlab{a}})Yusef-Zadeh, Cotton, Viti,
  Wardle, \& Royster}]{yusef-zadeh_widespread_2013}
Yusef-Zadeh, F., Cotton, W., Viti, S., Wardle, M., \& Royster, M.
  2013{\natexlab{a}}, The Astrophysical Journal Letters, 764, L19,
  \dodoi{10.1088/2041-8205/764/2/L19}

\bibitem[{Yusef-Zadeh {et~al.}(2013{\natexlab{b}})Yusef-Zadeh, Hewitt, Wardle,
  Tatischeff, Roberts, Cotton, Uchiyama, Nobukawa, Tsuru, Heinke, \&
  Royster}]{yusef-zadeh_interacting_2013}
Yusef-Zadeh, F., Hewitt, J.~W., Wardle, M., {et~al.} 2013{\natexlab{b}}, The
  Astrophysical Journal, 762, 33, \dodoi{10.1088/0004-637X/762/1/33}

\bibitem[{Zaleski {et~al.}(2013)Zaleski, Seifert, Steber, Muckle, Loomis,
  Corby, Martinez, Crabtree, Jewell, Hollis, Lovas, Vasquez, Nyiramahirwe,
  Sciortino, Johnson, McCarthy, Remijan, \& Pate}]{zaleski_detection_2013}
Zaleski, D.~P., Seifert, N.~A., Steber, A.~L., {et~al.} 2013, The Astrophysical
  Journal Letters, 765, L10, \dodoi{10.1088/2041-8205/765/1/L10}

\bibitem[{Zeng {et~al.}(2018)Zeng, Jiménez-Serra, Rivilla, Martín,
  Martín-Pintado, Requena-Torres, Armijos-Abendaño, Riquelme, \&
  Aladro}]{zeng_complex_2018}
Zeng, S., Jiménez-Serra, I., Rivilla, V.~M., {et~al.} 2018, Monthly Notices of
  the Royal Astronomical Society, 478, 2962, \dodoi{10.1093/mnras/sty1174}

\bibitem[{Zhao \& Truhlar(2004)}]{Zhao2004}
Zhao, Y., \& Truhlar, D.~G. 2004, Journal of Physical Chemistry A, 108, 6908,
  \dodoi{10.1021/jp048147q}

\bibitem[{Zheng \& Truhlar(2010)}]{zheng_kinetics_2010}
Zheng, J., \& Truhlar, D.~G. 2010, Physical Chemistry Chemical Physics, 12,
  7782, \dodoi{10.1039/B927504E}

\end{thebibliography}



\end{document}